\documentclass[a4paper,11pt]{article}
\usepackage{jheppub} 
\usepackage{tabularx,ragged2e}
\usepackage{orcidlink}
\usepackage{fontawesome5}
\usepackage{hyperref}
\newcolumntype{P}[1]{>{\RaggedRight\arraybackslash}p{#1}}

\usepackage{lineno}

\newcommand{\PS}{\ensuremath{\text{S}}}
\newcommand{\PB}{\ensuremath{\text{B}}}
\newcommand{\PX}{\ensuremath{\text{X}}}
\newcommand{\PZ}{\ensuremath{\text{Z}}}
\newcommand{\Ph}{\ensuremath{\text{h}}}
\newcommand{\PZD}{\ensuremath{\PZ_\text{D}}}

\newcommand{\PBzero}{\ensuremath{\text{B}^{0}}}

\newcommand{\PKstar}{\ensuremath{\text{K}^{(*)}}}

\newcommand{\mumu}{\ensuremath{\mu^{+}\mu^{-}}}
\newcommand{\ee}{\ensuremath{e^{+}e^{-}}}

\newcommand{\bb}{\ensuremath{b\bar{b}}}

\newcommand{\dd}{\ensuremath{d\bar{d}}}

\newcommand{\HtoXX}{\Ph\to\PX \PX}
\newcommand{\HtoSS}{\Ph\to\PS \PS}
\newcommand{\BtoKSS}{\PB\to\PKstar\PS\PS}
\newcommand{\BzerotoSS}{\PBzero\to\PS\PS}

\newcommand{\BhtoSS}{{\cal B}({\Ph\to\PS\PS})}
\newcommand{\BhtoZDZD}{{\cal B}({\Ph\to\PZD\PZD})}
\newcommand{\BhtoXX}{{\cal B}({\Ph\to\PX\PX})}

\newcommand{\BS}{{\cal B}{(\PS})}

\newcommand{\Bhtobb}{{\cal B}(\Ph \to \PX \PX \to 4b)}
\newcommand{\Bhtodd}{{\cal B}(\Ph \to \PX \PX \to 4d)}
\newcommand{\Bhtokk}{{\cal B}(\Ph \to \PX \PX \to 4K^{0})}
\newcommand{\Bhtopipi}{{\cal B}(\Ph \to \PX \PX \to 4\pi^{0})}
\newcommand{\Bhtotautau}{{\cal B}(\Ph \to \PX \PX \to 4\tau)}

\newcommand{\GeV}{\ensuremath{\text{GeV}}}
\newcommand{\TeV}{\ensuremath{\text{TeV}}}

\newcommand{\HtoZDZD}{\Ph\to\PZD\PZD}

\newcommand{\mS}{\ensuremath{m(\PS)}}
\newcommand{\mZD}{\ensuremath{m(\PZD)}}
\newcommand{\mX}{\ensuremath{m(\PX)}}
\newcommand{\mh}{\ensuremath{m(\Ph)}}

\title{\boldmath Dark scalar reinterpretation of searches for Higgs boson decays into long-lived particles at the LHC}

\author{Alberto Escalante del Valle~\orcidlink{0000-0002-9702-6359}}
\affiliation{Centro de Investigaciones Energ\'eticas Medioambientales y Tecnol\'ogicas (CIEMAT),\\
Avenida Complutense 40, 28040. Madrid, Spain}
\emailAdd{alberto.escalante@ciemat.es}

\abstract{
We present a reinterpretation of selected ATLAS and CMS searches for Higgs boson decays into long-lived particles (LLPs) within the scalar portal framework, where the standard model is extended by a light scalar that mixes with the Higgs boson. Using public \texttt{HEPData} results, we evaluate the sensitivity of several experimental strategies targeting different final states. The analysis shows how published searches probe complementary regions of the model parameter space, and it provides a comparative assessment of their coverage. The study also identifies the existing gaps in phase space coverage and highlights key areas for future improvement in LLP searches, which will be crucial to maximize the discovery potential for exotic decays of the Higgs boson into LLPs. Finally, we extrapolate the constraints to the integrated luminosity of the High-Luminosity LHC under different assumptions and compare them with the projected sensitivities of proposed dedicated LLP experiments at CERN, showing that these experiments will further extend the experimental reach independently of the assumed extrapolation.
}

\begin{document}
\maketitle
\flushbottom

\section{Introduction}
\label{sec:intro}
Searches for exotic decays of the standard model (SM) Higgs boson into long-lived particles (LLPs) have become increasingly important at the Large Hadron Collider (LHC), particularly motivated by beyond standard models (BSM) in which the Higgs boson acts as a portal to a hidden sector~\cite{Strassler:2006im}. Representative scenarios include decays to pairs of scalars, $\HtoSS$, or vectors, $\HtoZDZD$, which subsequently decay into standard model particles~\cite{Han:2007ae, Curtin:2014cca}; see also the recent review in Ref.~\cite{Cepeda:2021rql}. These searches target a wide range of final states, such as $\PZD\to\mu^{+}\mu^{-}$, which produces displaced $\mumu$, or $\PS\to \bb$, resulting in displaced dijets. Experimental collaborations designed searches exploiting various detector capabilities~\cite{Alimena:2019zri, Lee:2018pag}, including reconstruction of displaced vertices in the inner tracker or in the muon detectors, time-delayed signals in the calorimeters, or high-multiplicity hit clusters in the muon detectors. Different search strategies are sensitive to different LLP lifetimes, masses, and decay modes. This breadth of approaches reflects both the diversity of possible LLP signatures and the experimental challenges in their detection.

The landscape of LLP searches remains heterogeneous and difficult to navigate. Different analyses often adopt distinct benchmark models and assumptions for their interpretations, complicating direct comparisons across analyses and experiments. Many interpretations are often restricted to specific final states, not taking into account the interplay with other decay modes. Therefore, both assessing the overall coverage of the LLP parameter space and identifying gaps in the experimental program remain challenging.

In this paper we present a systematic reinterpretation of selected ATLAS and CMS searches for exotic decays of the Higgs boson into LLPs in various final states within the framework of the standard model extended by a scalar that mixes with the Higgs boson. This scenario, known as the scalar portal (BC5), is one of the eleven Benchmark Cases (BC1–BC11) proposed by the Physics Beyond Colliders (PBC) Study Group~\cite{Beacham:2019nyx}. Our reinterpretation builds on results from ATLAS and CMS obtained during LHC Run~2 (2015–2018), which collected approximately 140~fb$^{-1}$ of proton–proton collision data at a center-of-mass energy of $13~\TeV$, and the ongoing Run~3 (2022–2026), expected to deliver about 300~fb$^{-1}$ at $13.6~\TeV$. Using public results from \texttt{HEPData}~\cite{Maguire:2017ypu}, we evaluate the sensitivity of each relevant search and, for the first time, quantify the LHC constraints on the BC5 scenario. Our analysis highlights the complementary strengths of different experimental strategies and identifies regions of parameter space that remain weakly constrained. Furthermore, we compare the reach of current LHC searches with projections from proposed dedicated LLP experiments at CERN planned for the High-Luminosity LHC (HL-LHC) era (2030-2041), including SHiP~\cite{SHiP:2021nfo}, CODEX-b~\cite{CODEX-b:2019jve}, MATHUSLA~\cite{Curtin:2018mvb}, and ANUBIS~\cite{Bauer:2019vqk}, providing a broader perspective on the long term potential of LLP searches both within and beyond the main LHC detectors.

\section
{Model description: scalar portal}

The scalar portal is a minimal extension of the standard model Lagrangian, introducing one extra singlet field, $\PS$, and two types of couplings, $\mu$ and $\lambda$~\cite{OConnell:2006rsp}. Following the notation in~\cite{Beacham:2019nyx}, the additional terms in the Lagrangian are 
\begin{equation}
{\cal L}_{{\rm scalar}} = {\cal L}_{{\rm SM}} + {\cal L}_{{\rm DS}} - (\mu S + \lambda S^{2})H^{\dagger} H\,,
\label{eq:DarkHiggsLagrangian}
\end{equation}
where $H$ is the SM Higgs doublet. The extra scalar can serve as a mediator to a dark sector with a dark matter field, providing a potential interaction pathway described by ${\cal L}_{{\rm DS}}$.

The non-zero $\mu$ leads to the mixing of $h$ and $S$ states, where $h$ is the physical $125~\GeV$ Higgs boson. In the limit of small mixing angle, it can be written as
\begin{equation}
\theta = \frac{\mu v}{\mh^{2} - \mS^{2}}\,,
\label{eq:theta}
\end{equation}
where $v = 246~\GeV$, and $\mh$, $\mS$ are the masses of, respectively, the SM Higgs boson and the new scalar boson. The most general scenario, referred to as BC5 by the PBC Study group~\cite{Beacham:2019nyx}, considers $\lambda$, $\theta$, and $\mS$ as non-zero free parameters. In this framework, the $\PS$ boson is light, $\mS<\mh/2$, and its production is typically dominated by the quartic coupling $\lambda$. This leads to rare Higgs boson decays $\HtoSS$, relevant for ATLAS and CMS searches at the LHC, and to rare meson decays such as $\BtoKSS$ and $\BzerotoSS$, which are particularly well suited for intensity-frontier experiments and B-factories. This work focuses on a reinterpretation of selected LHC searches sensitive to the $\PS$ production via $\HtoSS$, assuming non-zero values of $\lambda$ that yield a branching fraction $\BhtoSS < 1\%$, consistent with the projected sensitivity at the HL-LHC to the branching fraction for the Higgs boson decaying into invisible final states, \(\mathcal{B}(\Ph \to \text{inv}) < 2.5\%\)~\cite{Cepeda:2019klc}. 

Through mixing with the Higgs boson, the light scalar inherits the Higgs boson couplings at that mass, suppressed by a factor of $\sin^{2}\theta$. Accordingly, its total width is given by
\begin{equation}
\Gamma(\PS) = \Gamma (\PS)_{\text{SM}} \sin^{2}\theta\,, 
\label{eq:Gamma}
\end{equation}
 where $\Gamma (\PS)_{\text{SM}}$ represents the width the scalar would have if it had Higgs-like couplings. For $\sin^{2} \theta$ values smaller than $\approx10^{-6}$, the light scalar mean proper decay length, $c\tau(\PS)$, is sufficiently large such that it can travel macroscopic distance before decaying, potentially yielding a distinctive LLP signature observable in the LHC detectors. This region of phase space is of particular phenomenological interest, as it can evade constraints from Higgs boson coupling measurements~\cite{CMS:2022dwd, ATLAS:2022vkf}, while simultaneously providing conditions consistent with a first-order electroweak phase transition~\cite{Carena:2022yvx}.

\begin{figure}[htbp]
\centering
\includegraphics[width=0.80\textwidth]{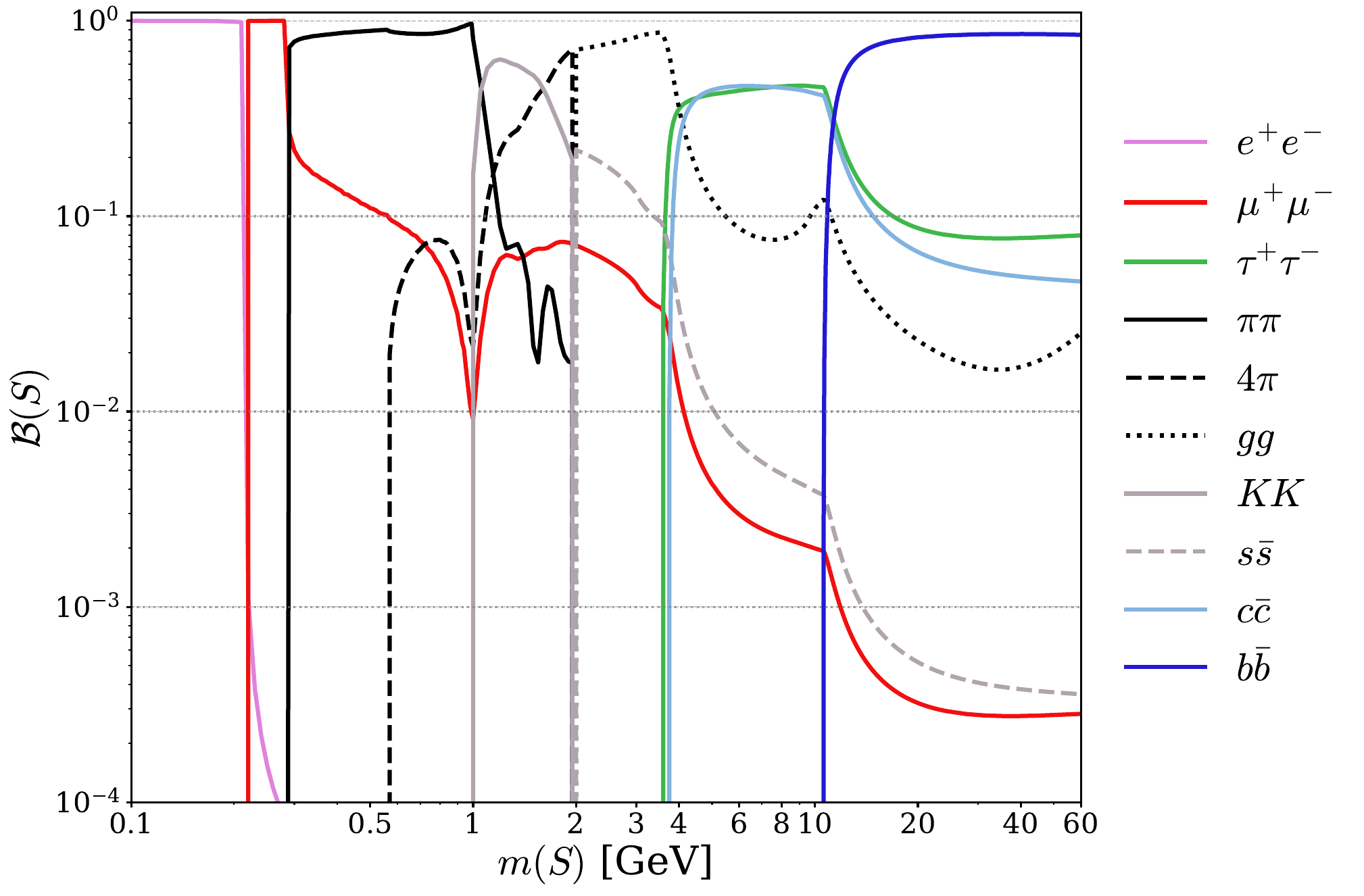}
\caption{Branching fractions of $\PS$ as a function of $\mS$ for various final states.}
\label{fig:S_branching_ratio}
\end{figure}

The branching fractions of the scalar decays in the various final states, $\BS$, depend strongly on $\mS$, reflecting its Yukawa-like coupling structure. For instance, the dominant decay mode for $\mS\gtrsim12~\GeV$ is $\PS\to b\bar{b}$, and decays to light leptons such as $\PS\to \mumu$ or $\PS\to \ee$ are highly suppressed. This work adopts the $\BS$ calculation from~\cite{Boiarska:2019jym}, using reformatted version that extends to higher scalar masses, shown in Fig.~\ref{fig:S_branching_ratio}. 

In Fig.~\ref{fig:S_branching_ratio} and throughout this work, we define ${\cal B}(\PS \to \pi\pi) = {\cal B}(\PS \to \pi^{\pm}\pi^{\mp}, \pi^{0}\pi^{0})$, ${\cal B}(\PS \to 4\pi) = {\cal B}(\PS \to \pi^{\pm}\pi^{\pm}\pi^{\mp}\pi^{\mp}, \pi^{\pm}\pi^{\mp}\pi^{0}\pi^{0})$, and ${\cal B}(\PS \to KK) = {\cal B}(\PS \to K^{\pm}K^{\mp}, K_{L}K_{S})$. It should be noted that in the range $\mS \in [0.7, 2]~\GeV$, significant theoretical uncertainties affect the calculation of scalar decays to hadrons~\cite{Boiarska:2019jym}.

\section{Reinterpretation methodology}

This section presents the selected LHC searches, the methodology for their reinterpretation within the BC5 model framework, and the main assumptions made in recasting the results. Section~\ref{subsec:leptonic} discusses leptonic final states, while Section~\ref{subsec:hadronic} focuses on hadronic final states.

\subsection{Leptonic searches}
\label{subsec:leptonic}
 Beginning with leptonic searches in the $\mu^{+}\mu^{-}$ final state, the experimental collaborations typically present their results as constraints on $\BhtoZDZD$, obtained in the framework of the Hidden Abelian Higgs Model (HAHM)~\cite{Curtin:2014cca}. The HAHM extends the standard model by introducing an additional dark gauge group, ${\cal U}(1)_{D}$, which mixes kinetically with the hypercharge gauge field and gives rise to a spin-1 mediator known as the dark photon, $\PZD$. This gauge symmetry is broken via a hidden sector Higgs mechanism, which introduces a scalar portal often referred to as the dark Higgs boson. The mixing between the dark Higgs boson and the SM Higgs boson enables the decay $\HtoZDZD$.  For simplicity, searches usually assume that the dark Higgs boson in the HAHM framework is sufficiently heavy to forbid its production in decays of the SM Higgs boson.
 
\begin{figure}[htbp]
\centering
\includegraphics[width=0.80\textwidth]{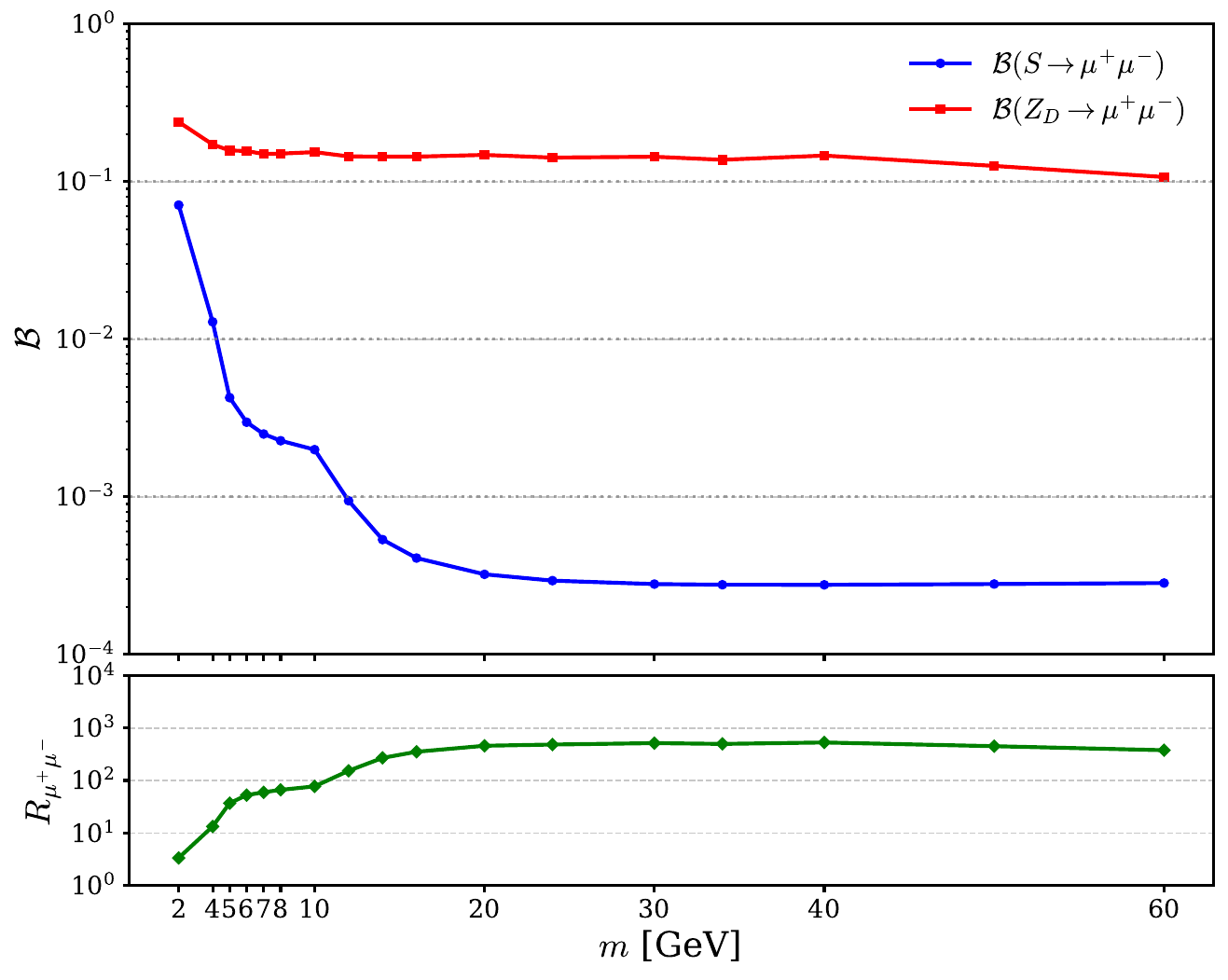}
\caption{Predicted ${\cal B}(\PZD \to \mu^{+}\mu^{-})$ in the HAHM (red), ${\cal B}(\PS \to \mu^{+}\mu^{-})$ in the scalar portal model (blue), and their ratio, $R_{\mu\mu}$ (green).}
\label{fig:ZD_D_Ratio}
\end{figure}

In the mass range considered in this study, $\mZD > 2~\GeV$, corresponding to the lowest $\mZD$ value for which numerical results are available, the predicted branching fraction ${\cal B}(\PZD\to\mumu)$ lies between $\approx30\%$ and $\approx10\%$, substantially larger than ${\cal B}(\PS\to\mumu)$, shown in Fig.~\ref{fig:S_branching_ratio}. As a result, the constraints on the Higgs boson branching fraction from displaced $\mumu$ searches would be significantly weaker if interpreted within the dark scalar framework. To translate the 95\% Confidence Level (CL) observed upper limits from $\BhtoZDZD$ into limits on $\BhtoSS$, we multiply the HAHM interpretation in \texttt{HEPData} using the rescaling factor

\begin{equation}
R_{\mu\mu} = {\cal B}(\PZD \to \mu^{+}\mu^{-})/{\cal B}(\PS \to \mu^{+}\mu^{-})\,, 
\label{eq:scale_mumu}
\end{equation}

where the predicted ${\cal B}(\PZD \to \mu^{+}\mu^{-})$, ${\cal B}(\PS \to \mu^{+}\mu^{-})$ and the rescaling factor $R_{\mu\mu}$ used in this study are shown in Fig.~\ref{fig:ZD_D_Ratio}.

Table~\ref{tab:Leptonic_finalstates} summarizes the selected $\mumu$ searches, the selected \texttt{HEPData} interpretation from the paper,  the available $\mZD$, and the rescaling factor applied to obtain the reinterpretation in the BC5 framework. To facilitate the assessment of the phase space coverage, only the most sensitive displaced $\mumu$ constraint is considered in  each region of the $\mZD, c\tau(\PZD)$ parameter space.

\begin{table}[ht]
    \centering
    \begin{tabular}{c c c P{5cm}}
         \hline
         Input from \texttt{HEPData} & $\mZD$ [$\GeV$]  & Factor & Experimental signature  \\
         \hline
         $\BhtoZDZD$ 
         & 60, 50, 40, 30, 20, 10
         & $R_{\mu\mu}$  
         & Displaced vertex in inner tracker and/or muon system, EXO-23-014~\cite{CMS:2024qxz} \\

         $\BhtoZDZD$ 
         & 12, 5, 2
         & $R_{\mu\mu}$ 
         & Displaced vertex in inner tracker with scouting data~\cite{CMS:2024zhe}, EXO-20-014~\cite{CMS:2021sch} \\
         \hline
         
    \end{tabular}
    \caption{Selected interpretation from \texttt{HEPData}, available $\mZD$, and rescaling applied to reinterpret displaced $\mumu$ searches in the BC5 framework.}
    \label{tab:Leptonic_finalstates}
\end{table}

The rescaling method based on Eq.~\ref{eq:scale_mumu} offers a significant advantage: it removes the need to re-run the full analyses performed by the experimental collaborations, which would otherwise require considerable time and computing resources. This approach, however, relies on several assumptions. First, it assumes similar signal efficiencies for scalar and vector LLPs pair produced in exotic decays of the Higgs boson, an assumption validated to within $\sim$20\% in Ref.~\cite{CMS:2024qxz}. Second, it assumes that differences between the two models in the fraction of events with two displaced $\mumu$ vertices have a negligible impact on the signal efficiency. Since this fraction is small in both models, the associated uncertainty is expected to be subdominant.
 
\subsection{Hadronic searches}
\label{subsec:hadronic}

For hadronic searches, LHC collaborations sometimes provide interpretations under the assumption of a $100\%$ branching fraction for each LLP into a single final state. This choice is widely used in CMS, while for ATLAS it is also common to use the branching fractions predicted by the extra scalar model. To distinguish an LLP with a fixed $100\%$ branching fraction from LLPs with the branching fractions predicted by the dark scalar model or HAHM, we denote this hypothetical LLP as $\PX$. Accordingly, there are LHC constraints on $\BhtoXX$ separated by $\PX$ decay mode: $b\bar{b}$, $d\bar{d}$, $\tau^{\pm}\tau^{\mp}$, $K^{\pm}K^{\mp}$, $K^{0}K^{0}$, $\pi^{\pm}\pi^{\mp}$, and $\pi^{0}\pi^{0}$, for representative LLP masses. Numerical results in \texttt{HEPData} are available for $\mX > 0.4~\GeV$, depending on the considered search.

While this signature driven approach is useful for designing analyses sensitive to various final states, it has limitations. As shown in Fig.~\ref{fig:S_branching_ratio}, the scalar branching fraction $\BS$ deviate from 100\% and vary strongly with $\mS$. Deriving constraints in a realistic set of models would ideally require signal efficiencies evaluated with the proper admixture of final states. However, providing results in both approaches ($\HtoXX$ and $\HtoSS$) is uncommon, as it demands substantial computing effort and time from the collaborations.
 
In this work, we adopt a conservative approach, meaning that it yields weaker constraints than a full re-analysis. Specifically, we take the published observed upper limits for a given final state, expressed in terms of $\BhtoXX$, and rescale them by the predicted fraction of scalar portal events that populate the same final state, or by events in other final states with comparable signal efficiencies. 

This strategy is applied to constraints by displaced dijets by ATLAS~\cite{ATLAS:2024qoo} and CMS~\cite{CMS:2024xzb} (relevant for $c\tau(\PS) \lesssim 0.1~\text{m}$), anomalous calorimeter energy deposits in ATLAS~\cite{ATLAS:2024ocv} (relevant for $c\tau(\PS) \approx 0.1~\text{m}$), and searches for high-multiplicity signals in the muon detectors by ATLAS~\cite{ATLAS:2018tup, ATLAS:2022gbw} and CMS~\cite{CMS:2024bvl} (relevant for $c\tau(\PS) \gtrsim 0.1~\text{m}$). The most stringent limits on $\BhtoXX$ from these searches, in the $\mX$ intervals $[55,60]~\GeV$, $[30,40]~\GeV$, and $[15,20]~\GeV$, are evaluated in the LHC LLP Summaries~\cite{ATL-PHYS-PUB-2025-002, CMS_Summary}. Since ATLAS and CMS do not provide interpretations at identical $\mX$ values, constraints are reported in these narrow intervals. We then rescale the numerical results by
\begin{equation}
1/{\cal B}(\PS \to b\bar{b})^{2}
\label{eq:scale_bb}
\end{equation}
to obtain LHC constraints on $\BhtoSS$ for $\mS > 15~\GeV$ in the BC5 framework. 

For $\mS < 15~\GeV$, the situation is more complex: the dominant $\PS$ decay mode depends on $\mS$, while experimental searches become increasingly challenging due to larger QCD and combinatorial backgrounds. At present, displaced vertex results by ATLAS and CMS do not place competitive constraints, and the strongest constraints arise from the recent CMS search for high-multiplicity hit clusters in the muon system~\cite{CMS:2024bvl}. In this case, the sensitivity is largely independent of the decay mode, with only a mild dependence on whether the shower is primarily hadronic ($b\bar{b}$, $d\bar{d}$, $K^{\pm}K^{\mp}$, $K^{0}K^{0}$, $\pi^{\pm}\pi^{\pm}$), electromagnetic ($\pi^{0}\pi^{0}$, $\gamma\gamma$, $e^{\pm}e^{\mp}$), or mixed ($\tau^{\pm}\tau^{\mp}$).

Motivated by this observation, and analogously to Eq.~\ref{eq:scale_bb}, we define rescaling factors that encode the main decay modes kinematically accessible for a given $\mS$, 
\begin{align}
R_{\pi\pi} &= 1/{\cal B}(\PS \to \pi\pi, 4\pi)^{2}\, , 
\label{eq:scale_pipi}\\[0.3cm]
R_{KK, 4\pi} &= 1/{\cal B}(\PS \to KK, 4\pi)^{2}\, , 
\label{eq:scale_kk}\\[0.3cm]
R_{gg, s\bar{s}} &= 1/{\cal B}(\PS \to gg, s\bar{s})^{2}\, , 
\label{eq:scale_gg}\\[0.3cm]
R_{c\bar{c}, \tau\tau} &= 1/{\cal B}(\PS \to gg, s\bar{s}, c\bar{c}, \tau_{\text{h}}\tau_{\text{h}})^{2}\, ,  
\label{eq:scale_cc}\\[0.3cm]
R_{b\bar{b}} &= 1/{\cal B}(\PS \to gg, s\bar{s}, c\bar{c},  \tau_{\text{h}}\tau_{\text{h}}, b\bar{b})^{2}\, , 
\label{eq:scale_qq}
\end{align}

where $\tau_{\text{h}}$ denotes a $\tau$ lepton that decays hadronically. The contribution from ${\cal B}(\PS \to u\bar{u})$ and ${\cal B}(\PS \to d\bar{d})$ are suppressed in the scalar model and negligible in Eqs.~\ref{eq:scale_pipi}--\ref{eq:scale_gg}. Table~\ref{tab:scale_factors_EXO-21-008} summarizes the rescaling factors and the main assumptions behind their definitions, given the signal efficiencies of EXO-21-008~\cite{CMS:2024bvl}.

\begin{table}[htbp]
    \centering
    \renewcommand{\arraystretch}{1.2}
    \begin{tabular}{c c c P{6.7cm}}
        \hline
        Input from \texttt{HEPData} & $\mX$ [GeV] & Factor & Assumptions for the limits in~\cite{CMS:2024bvl} \\
        \hline
    
        $\Bhtopipi$
        & 0.4, 1
        & $R_{\pi\pi}$ 
        & $\PX\to\pi^{0}\pi^{0}$ valid for $\pi^{\pm}\pi^{\mp}$ and $4\pi$ \\
        
        $\Bhtokk$
        & 1.5
        & $R_{KK, 4\pi}$ 
        & $\PX \to K^{0}K^{0}$ valid for $K^{\pm}K^{\mp}$ and $4\pi$\\
        
        $\Bhtodd$
        & 3
        & $R_{gg, s\bar{s}}$ 
        & $\PX \to \dd$ valid for $gg$ and $s\bar{s}$ \\
        
        $\Bhtodd$  
        & 7
        & $R_{c\bar{c}, \tau\tau}$
        & $\PX \to \dd$ valid for $gg$, $s\bar{s}$, $c\bar{c}$ and $\tau_{\text{h}}\tau_\text{h}$ \\
        
        $\Bhtobb$ 
        & 15
        & $R_{b\bar{b}}$ 
        & $\PX \to \bb$ valid for $gg$, $s\bar{s}$, $c\bar{c}$, $b\bar{b}$, and $\tau_{\text{h}}\tau_\text{h}$ \\
        
        \hline
        \end{tabular}
    \caption{Selected interpretation from \texttt{HEPData}, available $\mX$, and $R$ factors with associated assumptions used to reinterpret EXO-21-008~\cite{CMS:2024bvl} in the BC5 framework.}
    \label{tab:scale_factors_EXO-21-008}
\end{table}

The reinterpretation at $\mS = 7~\GeV$, which uses $R_{c\bar{c}, \tau\tau}$ and $\Bhtodd$, carries the largest uncertainty. At this mass, the predicted ${\cal B}(\PS\to c\bar{c})$ and ${\cal B}(\PS\to \tau^{+}\tau^{-})$ are comparable, but the constraints in Ref.~\cite{CMS:2024bvl} on $\Bhtodd$ are stronger than those on $\Bhtotautau$. The reduced efficiency arises mainly from leptonic $\tau$ decays such as $\tau \to \mu\nu\nu$, and to a lesser extent $\tau \to e\nu\nu$. Consequently, in Eq.~\ref{eq:scale_cc} we adopt the conservative choice of including only $\PS \to \tau_{h}\tau_{h}$ decays, which represent 42\% of all $\PS \to \tau\tau$ decays.

\begin{figure}[htbp]
\centering
\includegraphics[width=0.80\textwidth]{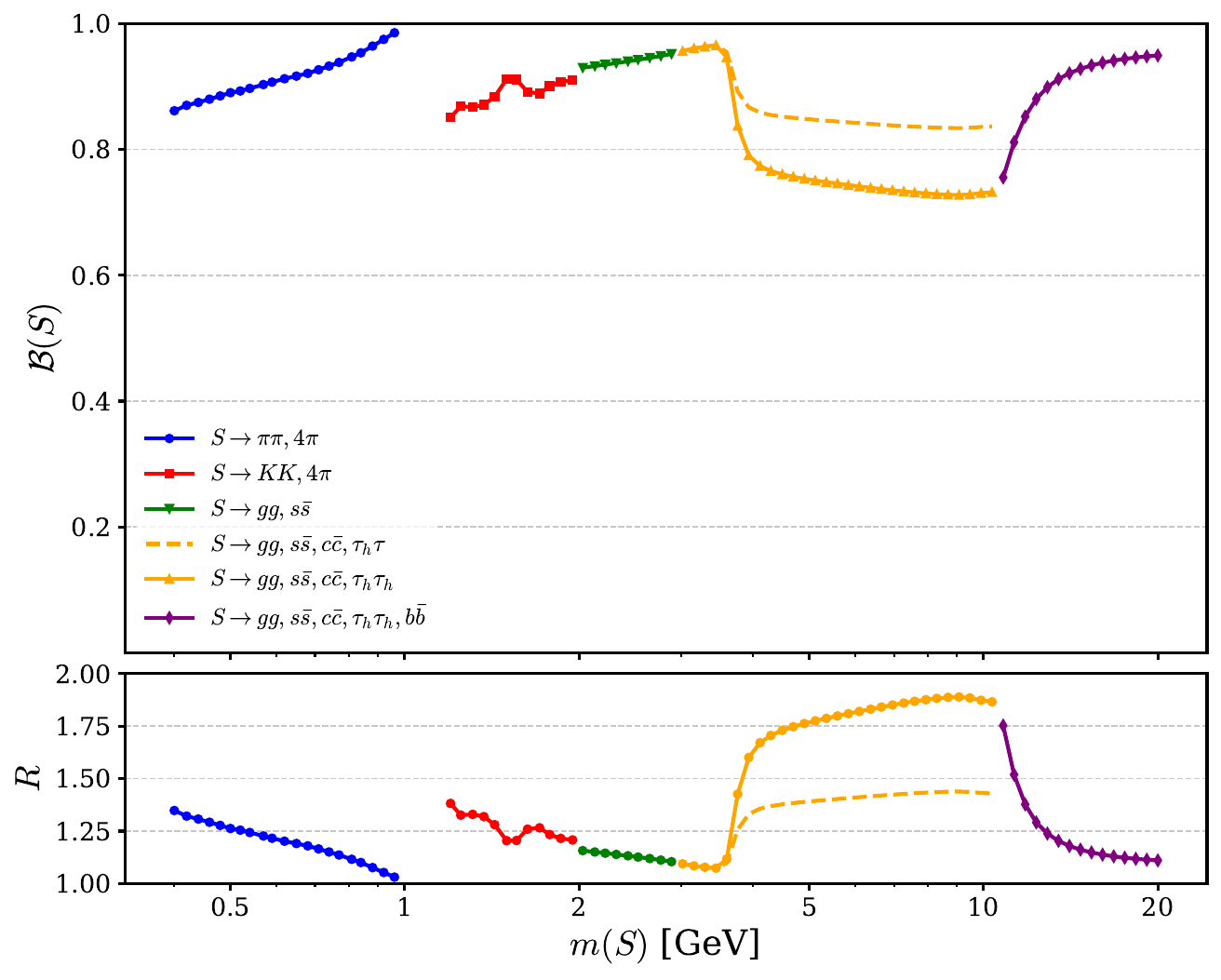}
\caption{The sum of the predicted $\BS$ in the considered hadronic final states in Eqs.~\ref{eq:scale_pipi}–\ref{eq:scale_qq} and the corresponding rescaling factors.}
\label{fig:R_had_ratio}
\end{figure}

The sum of the predicted $\BS$ in the considered final states of Eqs.~\ref{eq:scale_pipi}–\ref{eq:scale_qq}, shown in Fig.~\ref{fig:R_had_ratio}, exceeds 85\% for all considered $\mS$ values except at $\mS = 7~\GeV$, where it drops to 74\%. As a result, the $R$ factor is generally a small correction, and the published limits on $\PX$ are close to those for BC5. The largest rescaling factor is $R_{c\bar{c}, \tau\tau}$ (solid orange), reflecting the exclusion of $\tau$ leptonic decays.  A less conservative choice, including all $\tau$ modes with at least one hadronic decay, $\PS \to \tau_{h}\tau$ (dashed orange), would strengthen the limits at $\mS = 7~\GeV$ by 28\%.

The difference in the rescaling factor of Eq.~\ref{eq:scale_mumu} compared to Eqs.~\ref{eq:scale_bb}–\ref{eq:scale_qq} comes from how signal efficiencies are evaluated by the collaborations in the considered searches. For instance, in displaced $\mumu$ searches, efficiencies are derived in the HAHM model, where $\HtoZDZD$ decays typically yield a single displaced dimuon vertex. By contrast, in hadronic searches the $\PX$ benchmark model contains two displaced vertices.

\section{Results}

This section presents the results of the reinterpretation in the BC5 framework. Section~\ref{subsec:lhc} discusses the current constraints from selected ATLAS and CMS publications, while Section~\ref{subsec:hl-lhc} addresses their extrapolation to HL-LHC conditions and compares them with the projected sensitivities of dedicated LLP experiments at CERN.

\subsection{LHC constraints}
\label{subsec:lhc}

\begin{figure}[htbp]
  \centering
  \includegraphics[width=0.49\textwidth]{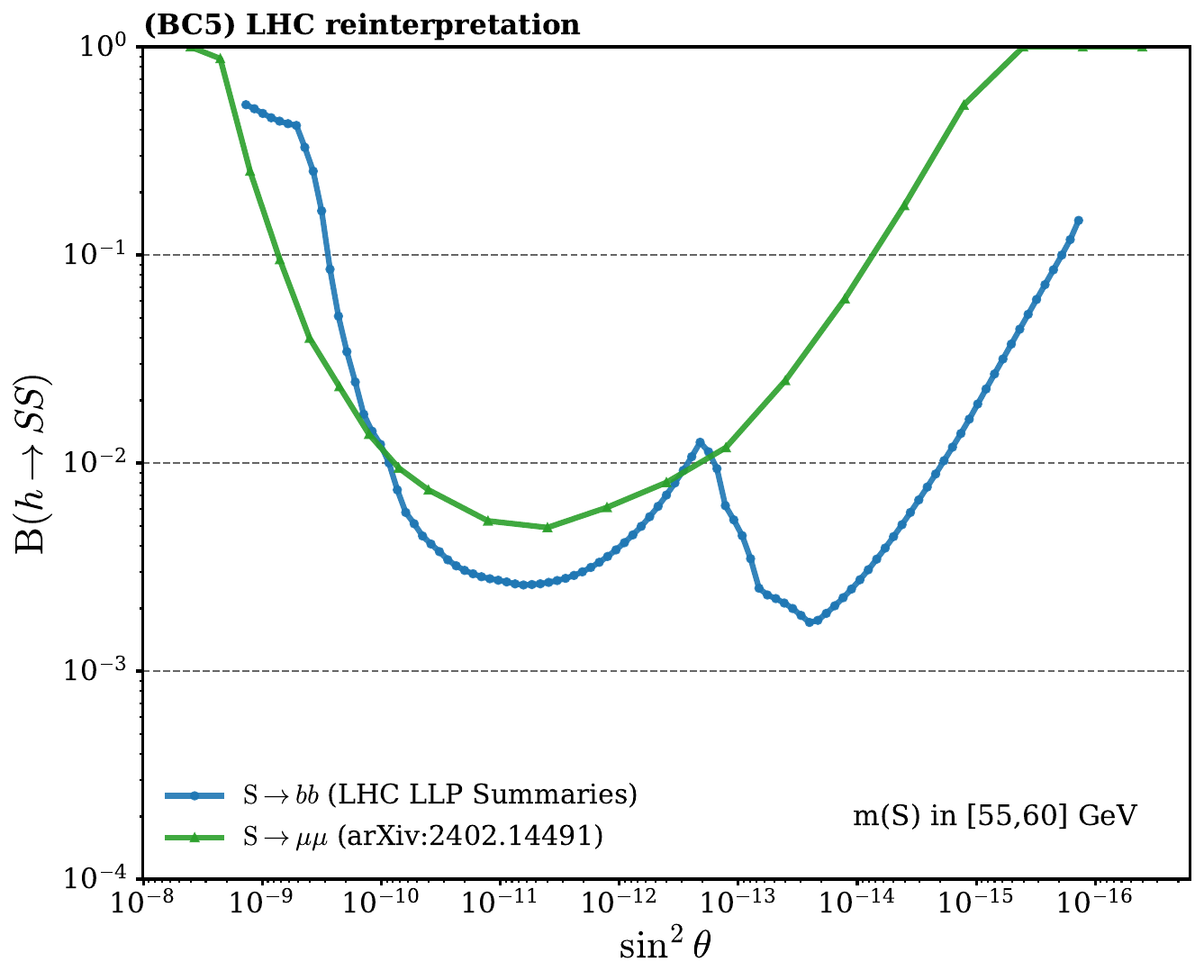}
  \includegraphics[width=0.49\textwidth]{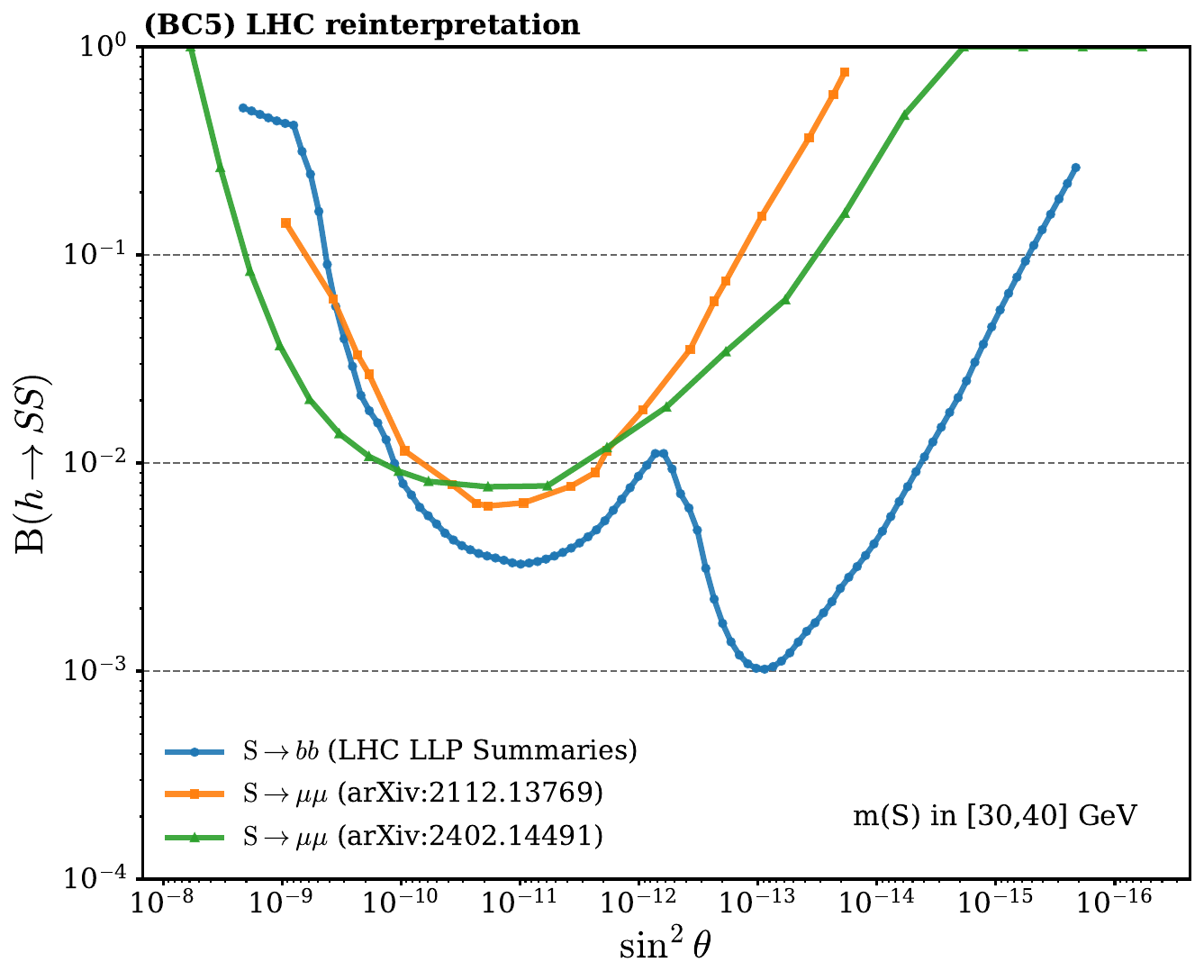}

  \includegraphics[width=0.49\textwidth]{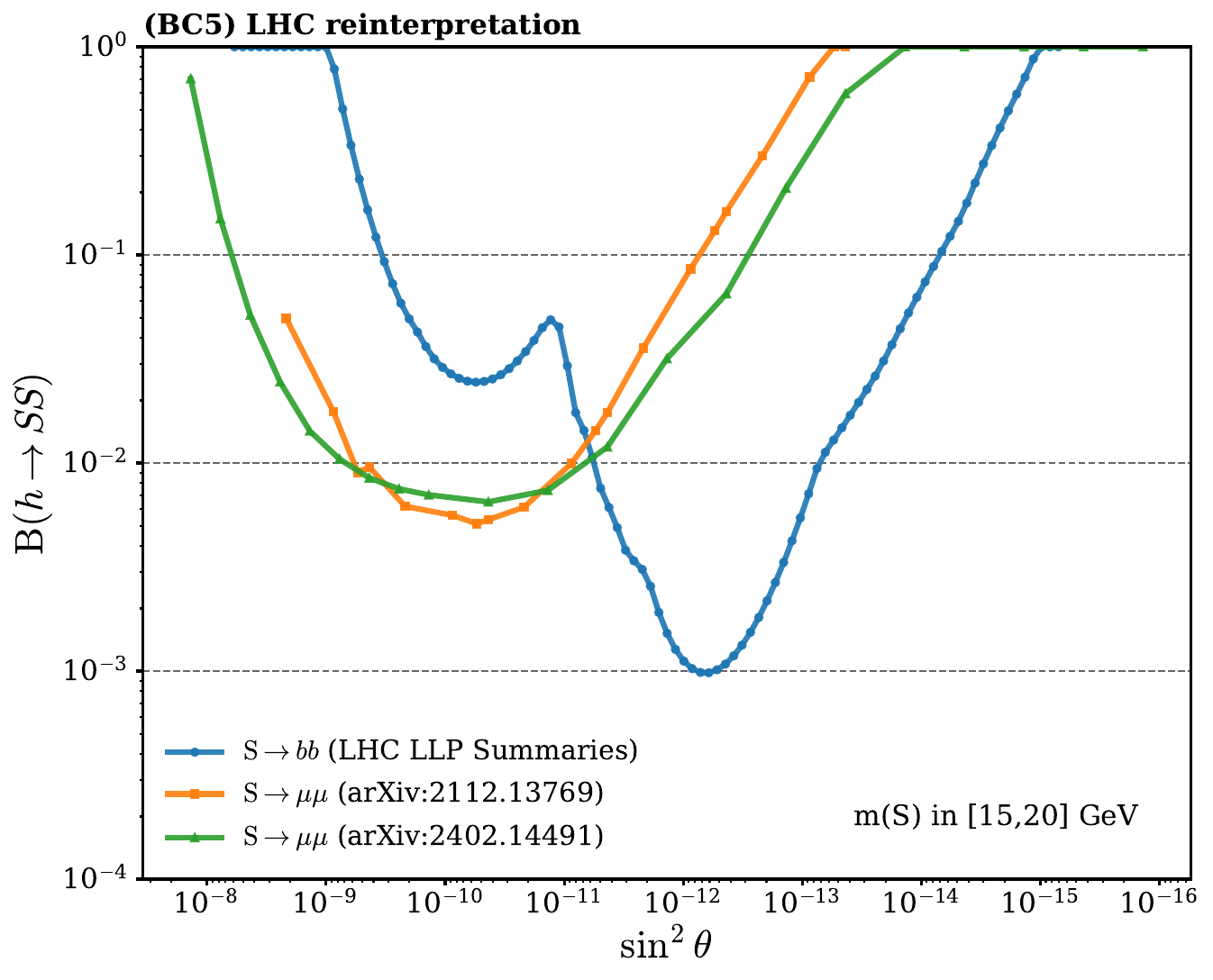}
  \includegraphics[width=0.50\textwidth]{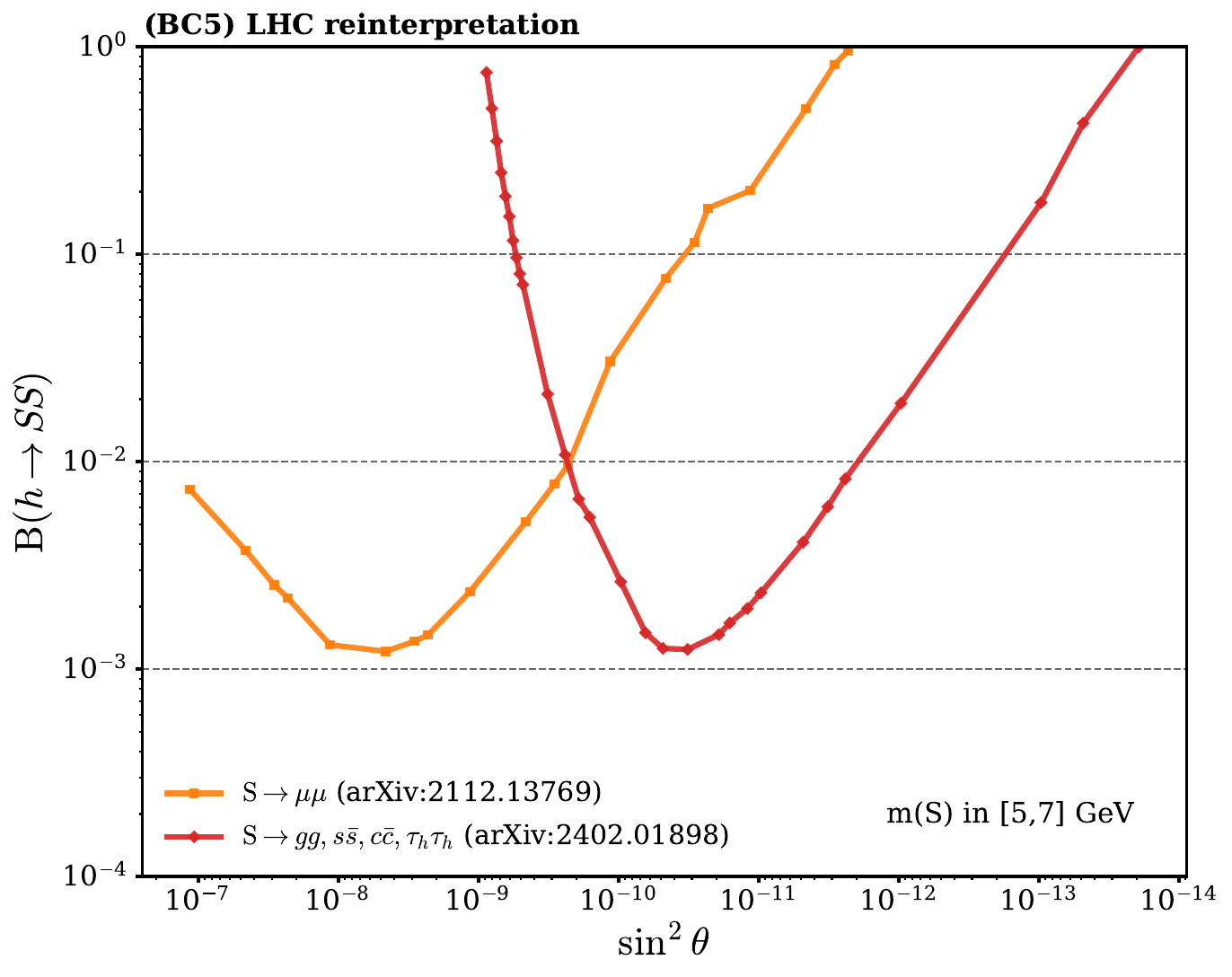}

  \includegraphics[width=0.49\textwidth]{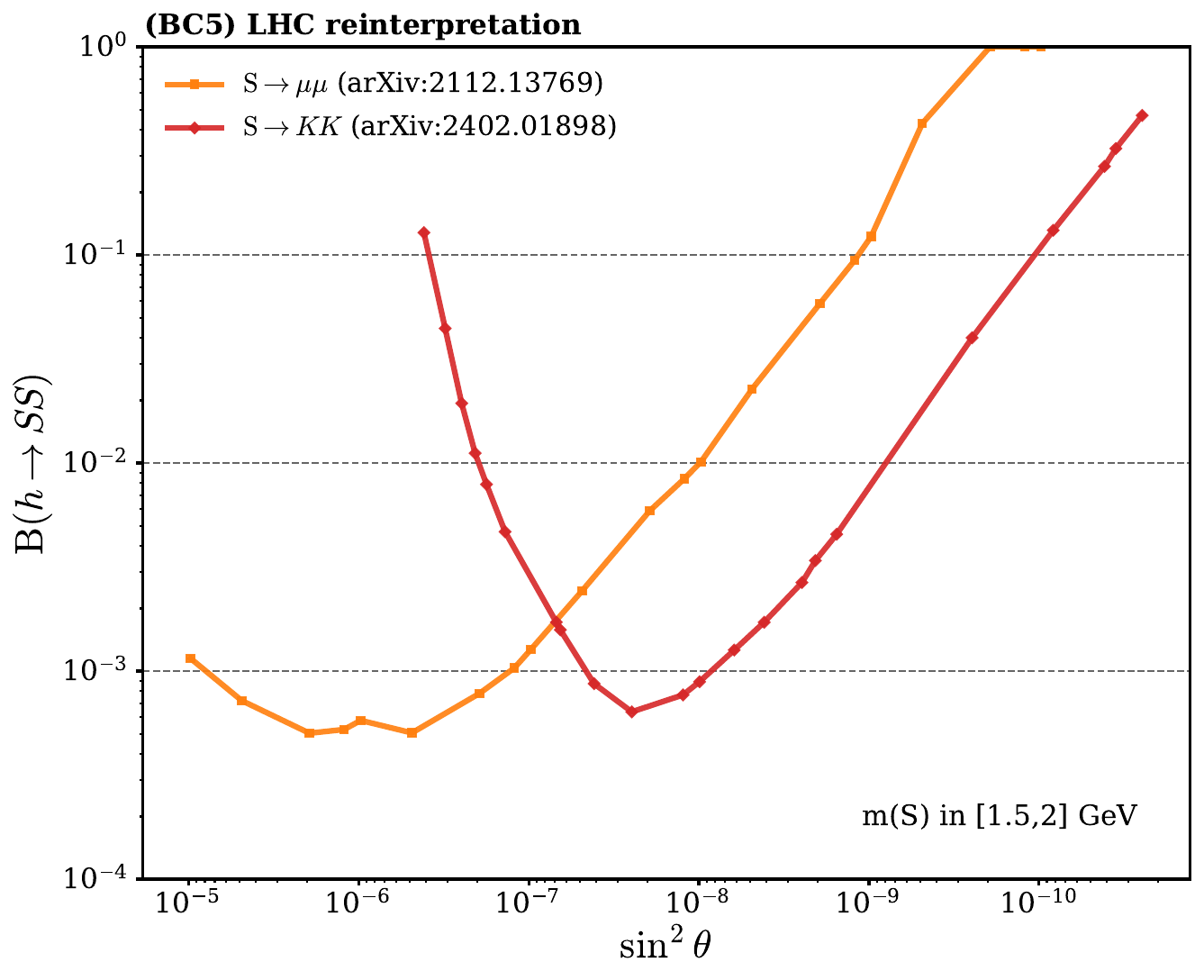}
  \includegraphics[width=0.50\textwidth]{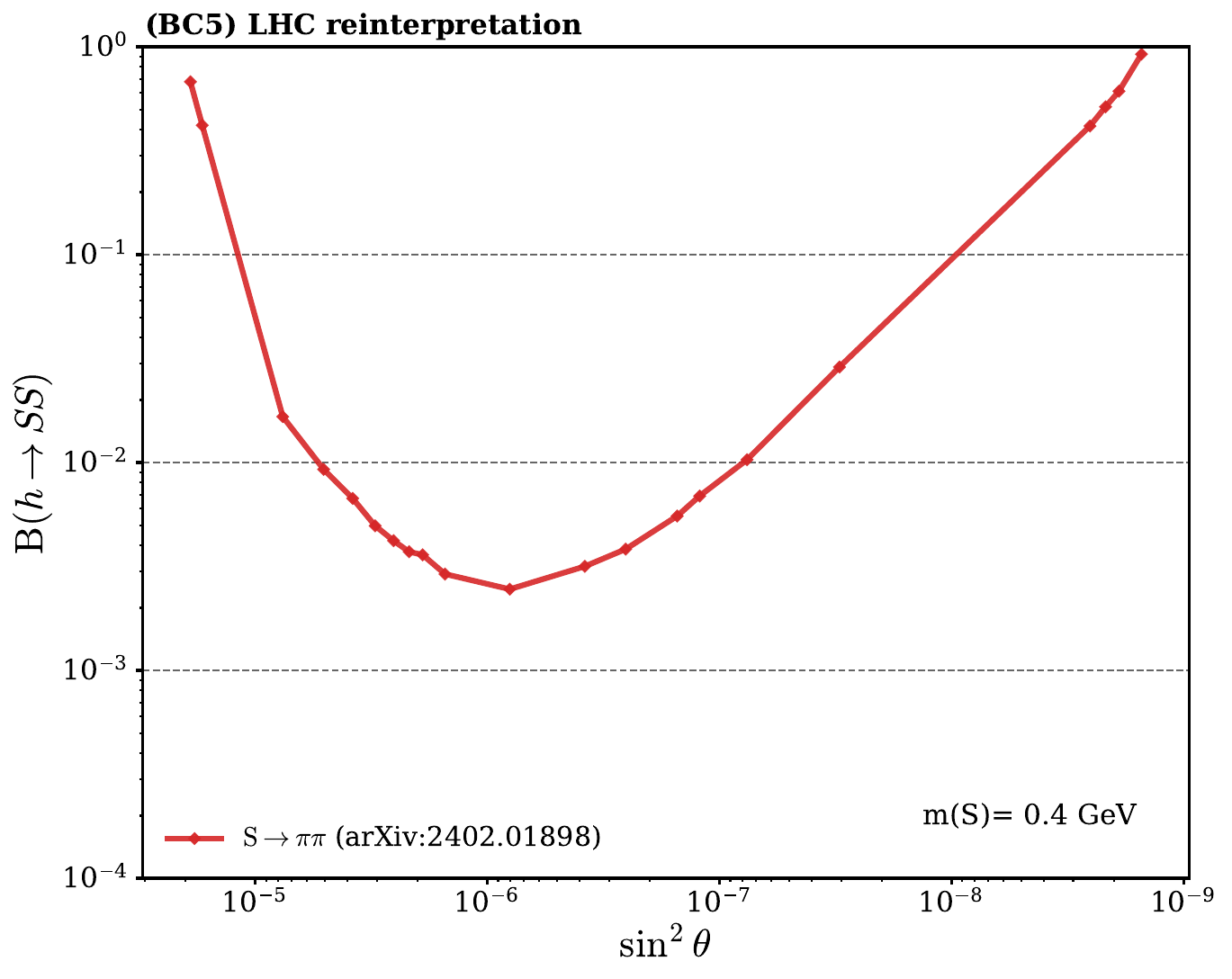}

  \caption{95\% CL observed upper limits on $\BhtoSS$ as a function of $\sin^{2}\theta$ in the BC5 scalar portal model, for $\mS$ ranging from [55,60]~$\GeV$ (upper left) to 0.4~$\GeV$ (lower right), obtained from the reinterpretation of selected LHC LLP searches.}

  \label{fig:Upper_limits}
\end{figure}

In Fig.~\ref{fig:Upper_limits}, we show 95\% CL observed upper limits from the reinterpretation of leptonic channels: EXO-23-014~\cite{CMS:2024qxz} (green) and EXO-20-014~\cite{CMS:2021sch} (orange), together with hadronic ones: LHC LLP Summaries which include results from~\cite{ATLAS:2024qoo, CMS:2024xzb, ATLAS:2024ocv, ATLAS:2018tup, ATLAS:2022gbw, CMS:2024bvl} (blue) and EXO-21-008~\cite{CMS:2024bvl} (red), as a function of $\mS$. Details of the reinterpretation are provided in Tables~\ref{tab:Leptonic_finalstates}, 
\ref{tab:scale_factors_EXO-21-008}.

Starting at the largest masses, $\mS\in[55,60]~\GeV$, the best constraints in most of the probed $\sin^{2}\theta$ range are provided by hadronic searches, except for the transition region in $\sin^{2}\theta$ from $\approx10^{-13}$ to $\approx10^{-12}$ where a gap in experimental coverage remains between displaced vertices~\cite{CMS:2024xzb}, and high-multiplicity showers in the muon detectors~\cite{CMS:2024bvl}. In this region, searches for displaced $\mumu$ vertices using exclusively the muon system~\cite{CMS:2024qxz} provide competitive sensitivity. At the shortest lifetimes, $\sin^{2}\theta \gtrsim 10^{-10}$, displaced dimuons, this time exploiting the impact parameter resolution of the inner detector, extend the coverage over displaced vertices. 

At slightly lower masses, $\mS \in [30,40]~\GeV$, the experimental coverage is similar, again highlighting the additional reach at short lifetimes from displaced $\mumu$ searches in the inner tracker. Constraints from high-multiplicity showers in the muon detectors remain the most powerful probe for $\sin^{2}\theta \lesssim 10^{-12}$.
 
In the mass region just above the $\Upsilon$ resonances, $\mS\in[15,20]~\GeV$, the performance of displaced vertex analyses deteriorates, which leads to displaced $\mumu$ results substantially extending the reach in $\sin^{2}\theta$, improving over displaced vertex searches and high-multiplicity showers in the muon detectors at short lifetimes for $\sin^{2} \theta \gtrsim~10^{-11}$.

At $\mS \in [5,7]~\GeV$, between the $J/\psi$ and $\Upsilon$ resonances, where $\PS$ predominantly decays to $c\bar{c}$ and $\tau\tau$, displaced vertex searches lose further sensitivity due to the increased QCD and combinatorial background. In this regime, displaced dimuon results based on the scouting dataset~\cite{CMS:2021sch} provide essential coverage, enabling exploration down to $\sin^{2}\theta \gtrsim 10^{-9}$, despite the significantly lower predicted ${\cal B}(\PS \to \mumu)$ compared to other final states. A gap appears around $\sin^{2}\theta \approx 5 \times 10^{-9}$, corresponding to the transition region towards muon detector shower signatures.

For lighter scalar masses, $\mS \in [1.5,2]~\GeV$, below the $J/\psi$ resonance, the decay $\PS \to KK$ becomes sizable. In this region, there is again complementarity between $\mumu$ results and hadronic decays in the muon detectors. The sensitivity is the strongest across all $\mS$ values, reaching $\BhtoSS \lesssim 10^{-3}$ for $\sin^{2}\theta$ between $\approx 10^{-7}$ and $\approx 10^{-5}$ and in a narrow window around $\approx 10^{-8}$. Compared to higher masses, displaced $\mumu$ searches play a particularly important role due to the increase in $\mathcal{B}(\PS \to \mumu)$. 

At the lowest probed masses, $\mS = 0.4~\GeV$, the decay $\PS \to \pi\pi$ dominates, and the CMS search for electromagnetic showers in the muon detectors provides the strongest constraints, reaching $\BhtoSS < 10^{-2}$ for $\sin^{2}\theta$ between $\approx 3 \cdot 10^{-7}$  and $\approx 5\cdot 10^{-5}$. Constraints from searches for displaced $\mumu$ resonances are weak (not shown) for $\mS$ values near SM meson resonances at 0.46, 0.55, 0.78, and 1.02~$\GeV$, corresponding to $K_S$, $\eta$, $\rho/\omega$, and $\phi(1020)$, respectively. These resonances could mimic a signal and are masked in Ref.~\cite{CMS:2021sch}. 

\begin{figure}[htbp]
  \centering
  \includegraphics[width=0.80\textwidth]{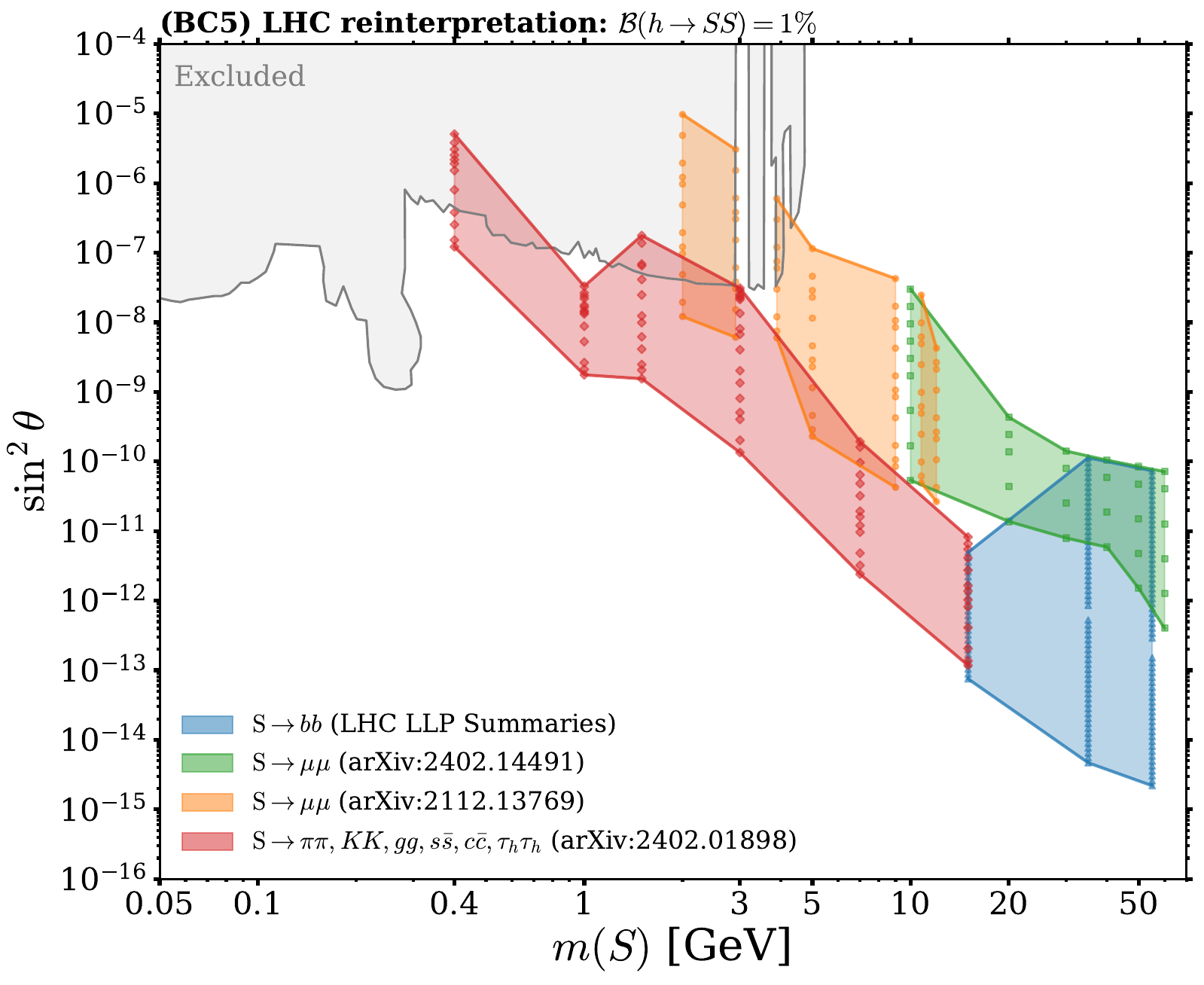}\\
  \caption{95\% CL observed upper limits in the $\{\mS, \sin^{2}\theta\}$ plane, assuming $\BhtoSS = 1\%$, in the BC5 scalar portal model. Shown are the reinterpretation of LHC LLP Summaries including Refs.~\cite{ATLAS:2024qoo, CMS:2024xzb, ATLAS:2024ocv, ATLAS:2018tup, ATLAS:2022gbw, CMS:2024bvl} (blue), EXO-23-014~\cite{CMS:2024qxz} (green), EXO-20-014~\cite{CMS:2021sch} (orange), and EXO-21-008~\cite{CMS:2024bvl} (red). The results are compared with current constraints from Ref.~\cite{PBC:2025sny} (gray), which include results from E949, NA62, MicroBooNE, KOTO, ICARUS, LHCb, and Belle~II as well as reinterpretations of PS191, CHARM and LSND.}
  \label{fig:Sumamry_Br_0p01}
\end{figure}

In Fig.~\ref{fig:Sumamry_Br_0p01}, we translate the 95\% CL exclusions from Fig.~\ref{fig:Upper_limits} at $\BhtoSS = 1\%$ into 2D contours in the $\{\mS, \sin^{2}\theta\}$ plane. The reinterpretation is performed at discrete $\mS, \sin^{2}\theta$ points, indicated by markers, corresponding to the $\mS$ values listed in Tables~\ref{tab:Leptonic_finalstates} and \ref{tab:scale_factors_EXO-21-008}. The values obtained at these discrete points were connected by straight lines in a log-log plane. To construct continuous 2D contours for displaced $\mumu$ searches in the $\mS \in [2, 12]~\GeV$ range, additional exclusion points at $\mS = 2.9, 3.9, 8.9,$ and $10.8~\GeV$ were estimated by extrapolating from the nearest available $\mS$ interpretations. These estimates should be regarded as approximate but are essential to highlight the coverage gaps caused by the masking of $J/\psi$, and $\Upsilon$ resonances. 

The region near the threshold, $\mS\approx 62.5~\GeV$, is covered by all considered ATLAS and CMS searches. However, no extrapolation from nearby $\mS$ values was attempted. To first approximation, the constraints obtained at $\mS\in[55,60]~\GeV$ may be taken as valid up to $\mS\approx 62.5\GeV$, though with large uncertainties.

The results are compared with current constraints (gray) from Ref.~\cite{PBC:2025sny}, which include results from E949, NA62, MicroBooNE, KOTO, ICARUS, LHCb, and Belle~II, as well as reinterpretations of PS191, CHARM, and LSND. The comparison highlights how the phase space probed by ATLAS and CMS complements that covered by other experiments. Moreover, the complementarity among individual contours (orange, green, red, and blue) shows that hadronic and leptonic searches together are needed to systematically explore the $\{\mS, \sin^{2}\theta\}$ plane beyond existing bounds. 

The uncovered region of large mixing in Fig.~\ref{fig:Sumamry_Br_0p01}, $\sin^{2}\theta \gtrsim 10^{-6}$ for $\mS \gtrsim 5~\GeV$, is constrained by Higgs boson coupling measurements and by searches for exotic decays of the Higgs boson with promptly decaying $\PS$ (not shown). While important, this part of $\{\mS, \sin^{2}\theta\}$ plane is outside the scope of the present study, which focuses on the reinterpretation of LLP searches.

\begin{figure}[htbp]
  \centering
  \includegraphics[width=0.80\textwidth]{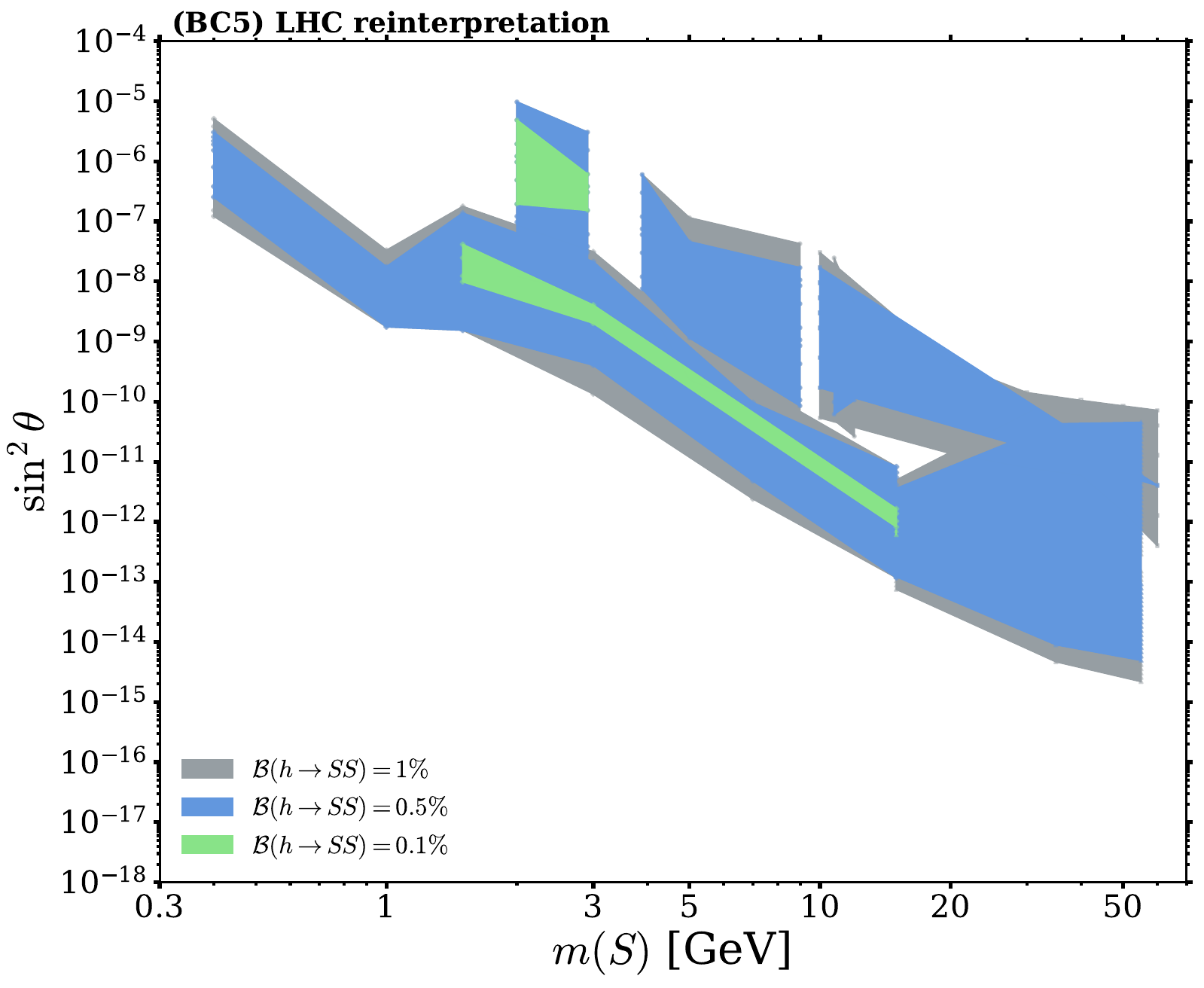}\\
  \caption{95\% CL observed upper limits in the $\{\mS, \sin^{2}\theta\}$ plane for $\BhtoSS$: 1\% (gray), 0.5\% (blue), and 0.1\% (green), in the BC5 scalar portal model, considering all selected LHC searches.}
  \label{fig:Sumamry_byBr}
\end{figure}

The excluded regions in Fig.~\ref{fig:Sumamry_Br_0p01} assume $\BhtoSS = 1\%$, a common benchmark in the literature, while well motivated BSM models also predict smaller values~\cite{Carena:2022yvx}. Fig.~\ref{fig:Sumamry_byBr} shows the 95\% CL excluded regions in the $\{\mS, \sin^{2}\theta\}$ plane for representative values of $\BhtoSS$, demonstrating that at $\BhtoSS = 0.1\%$ (green) much of the parameter space remains unconstrained. This highlights the need for further improvements in Run~3 and at the HL-LHC to maximize discovery potential and avoid gaps in coverage.

While this manuscript was in preparation, CMS updated the displaced $\mumu$ constraints for $\mS < 5~\GeV$ using the first 64.2~fb$^{-1}$ of Run~3 scouting data~\cite{CMS-PAS-EXO-24-016}, and released a new search targeting $\HtoSS$ with mixed $\mumu$ and hadronic decays  ($\pi^{\pm}\pi^{\mp}$ or $K^{\pm}K^{\mp}$)~\cite{CMS-PAS-EXO-24-034}, reaching $\BhtoSS \approx 0.1\%$ at $\mS = 0.6~\GeV$ for $c\tau(\PS) = 5$~mm. ATLAS also updated its displaced vertex search in the muon detectors~\cite{ATLAS:2025pak}. The addition of these new results is not expected to significantly change any of the conclusions of this paper. 

\subsection{HL-LHC projections}
\label{subsec:hl-lhc}

The HL-LHC is expected to deliver about 3~ab$^{-1}$ of data in ATLAS and CMS by 2041. In addition to increasing the dataset size by about an order of magnitude, this dataset will be collected with upgraded detectors. These upgrades are expected to significantly enhance the BSM discovery potential, particularly for LLP searches. Key improvements include extended detector acceptance, new high-granularity calorimeters, dedicated timing detectors, and more advanced data acquisition systems enabling improved trigger capabilities.

A key limitation of current searches is that their signal efficiencies are often limited by trigger performance, as the existing algorithms were not originally designed to capture LLP signals. Trigger development for LLPs is an active research area in CMS, ATLAS, and LHCb~\cite{CMS:2025ffw, ATLAS:2024xna, Vagnoni:2025qfv}, therefore the discovery potential for Run 3, and especially during the HL-LHC, is expected to improve substantially. However, no realistic HL-LHC projections based on Run~2 searches are currently available, and a simple scaling of current constraints by $\sqrt{\cal L}$, which accounts only for the increase in integrated luminosity, generally underestimates the true reach. In fact, early Run 3 searches by ATLAS~\cite{ATLAS:2024vnc} and CMS~\cite{CMS:2024qxz, CMS:2024xzb} have already demonstrated increased sensitivity significantly exceeding the scaling by $\cal L$ when compared with the predecessor Run~2 publications, thanks to improved trigger algorithms and refined analysis techniques.

\begin{figure}[htbp]
  \centering
  \includegraphics[width=0.80\textwidth]{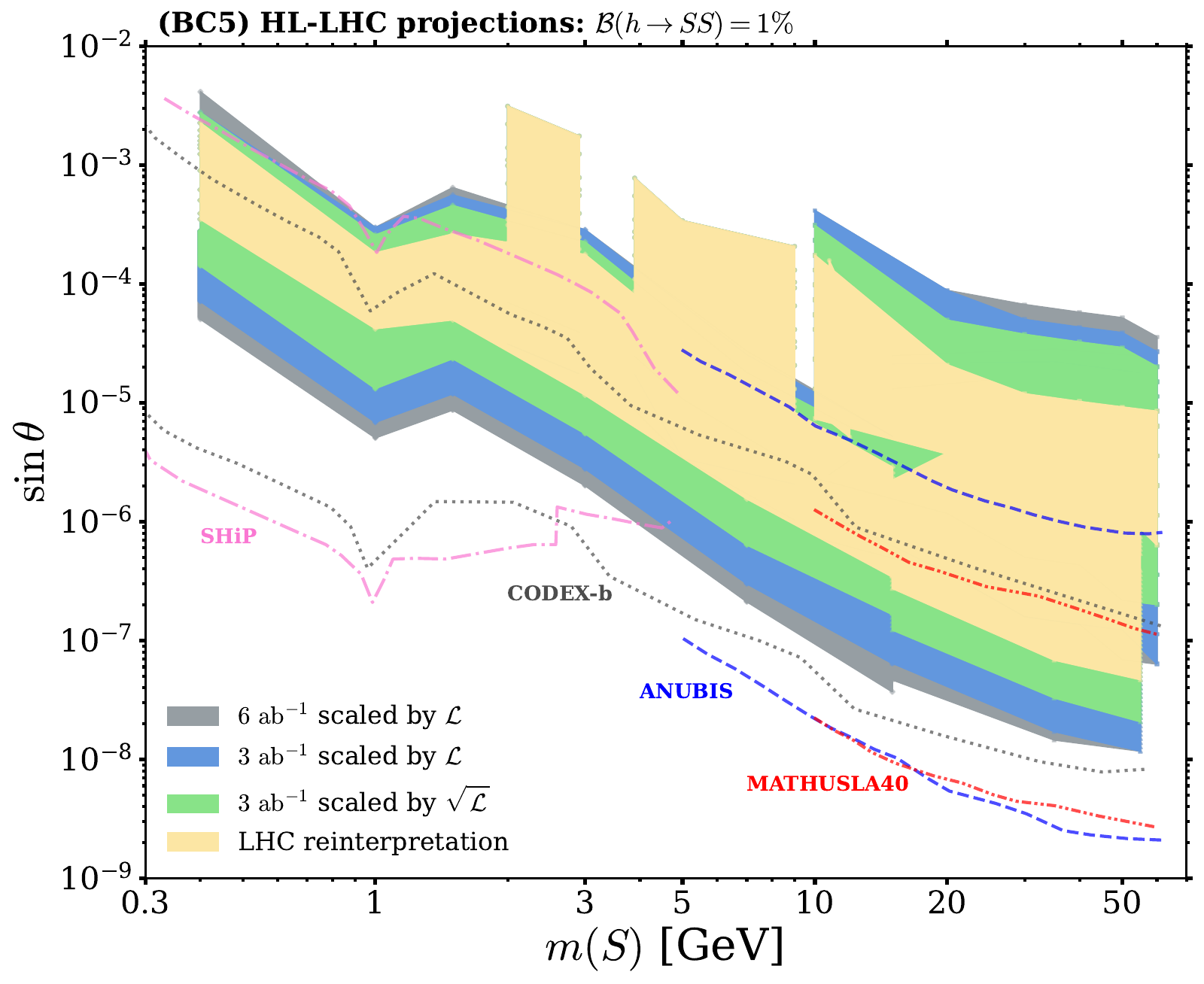}\\
  \caption{95\% CL observed upper limits in the $\{\mS, \sin \theta\}$ plane, assuming $\BhtoSS = 1\%$, in the BC5 scalar portal model. Shown are the reinterpretation of selected LHC searches (yellow) and their HL-LHC extrapolations: 3~ab$^{-1}$ scaled by $\sqrt{\cal L}$ (green), 3~ab$^{-1}$ scaled by ${\cal L}$ (blue), and 6~ab$^{-1}$ scaled by ${\cal L}$ (gray). The projections are compared with sensitivity estimates from SHiP, CODEX-b, MATHUSLA, and ANUBIS from Ref.~\cite{PBC:2025sny}.}
  \label{fig:HL-LHC_0.01}
\end{figure}

In Fig.~\ref{fig:HL-LHC_0.01} we compute 95\% CL upper limits assuming three types of HL-LHC extrapolations from our BC5 constraints with $\BhtoSS = 1\%$ (yellow):

\begin{itemize}
    \item 3~ab$^{-1}$ scaled by $\sqrt{\cal{L}}$ (green): current 95\% CL observed upper limits are scaled by $\sqrt{140/3000}$. This corresponds to a conservative scenario in which backgrounds remain and no improvements in signal efficiency are achieved at the HL-LHC.
    \item 3~ab$^{-1}$ (blue) and 6~ab$^{-1}$ (gray) scaled by $\cal{L}$: current 95\% CL observed upper limits are scaled by $140/3000$ and $140/6000$, respectively. The 3~ab$^{-1}$ case correspond to an optimistic scenario where backgrounds can be reduced to negligible levels thanks to HL-LHC upgrades and new developments in analysis techniques. The 6~ab$^{-1}$ case corresponds to either doubled signal efficiencies (e.g., from improved trigger algorithms), or a combination of ATLAS and CMS datasets.
\end{itemize}

In each HL-LHC projection, all upper limits are assumed to scale equally. In practice, however, some search strategies offer greater potential for R\&D than others, which should be kept in mind when interpreting these projections. Since LHC LLP summaries are available only for 95\% CL observed limits, all extrapolations are based on observed limits. Because of the limited granularity of interpretations at the smallest values of $\sin \theta$ (e.g see Fig.~\ref{fig:Upper_limits} for $\mS < 10~\GeV$), additional points were obtained by fitting the constraints with a linear regression in log-log  plane to produce smoother contours in Fig.~\ref{fig:HL-LHC_0.01}. These HL-LHC projections are compared with the expected sensitivity estimates from Ref.~\cite{PBC:2025sny} for dedicated LLP detectors at CERN. Following the format in Ref.~\cite{PBC:2025sny}, the comparison is performed in the $\{\mS, \sin \theta\}$ plane instead of $\{\mS, \sin^{2}\theta\}$.

For $\mS\gtrsim 2.5~\GeV$, the proposed CODEX-b, MATHUSLA, and ANUBIS experiments are expected to improve the HL-LHC constraints on $\sin \theta$ by approximately one order of magnitude, depending on the LLP experiment, $\mS$, and assumed extrapolation. In the region $0.4 \lesssim \mS \lesssim 2.5~\GeV$, the most stringent projections come from the SHiP experiment, already approved, which aims to extend the constraint on $\sin \theta $ by one to two orders of magnitude, depending on $\mS$, and assumed HL-LHC extrapolation. The SHiP experiment targets the $\PS$ production in meson decays, which makes the experiment also sensitive to the BC4 scenario, where $\lambda = 0$ in Eq.~\ref{eq:DarkHiggsLagrangian}. Overall, the study highlights that all proposed LLP experiments will explore new regions in the $\{\mS, \sin \theta\}$ plane, independent of the assumed HL-LHC extrapolation.

\begin{figure}[htbp]
  \centering
  \includegraphics[width=0.80\textwidth]{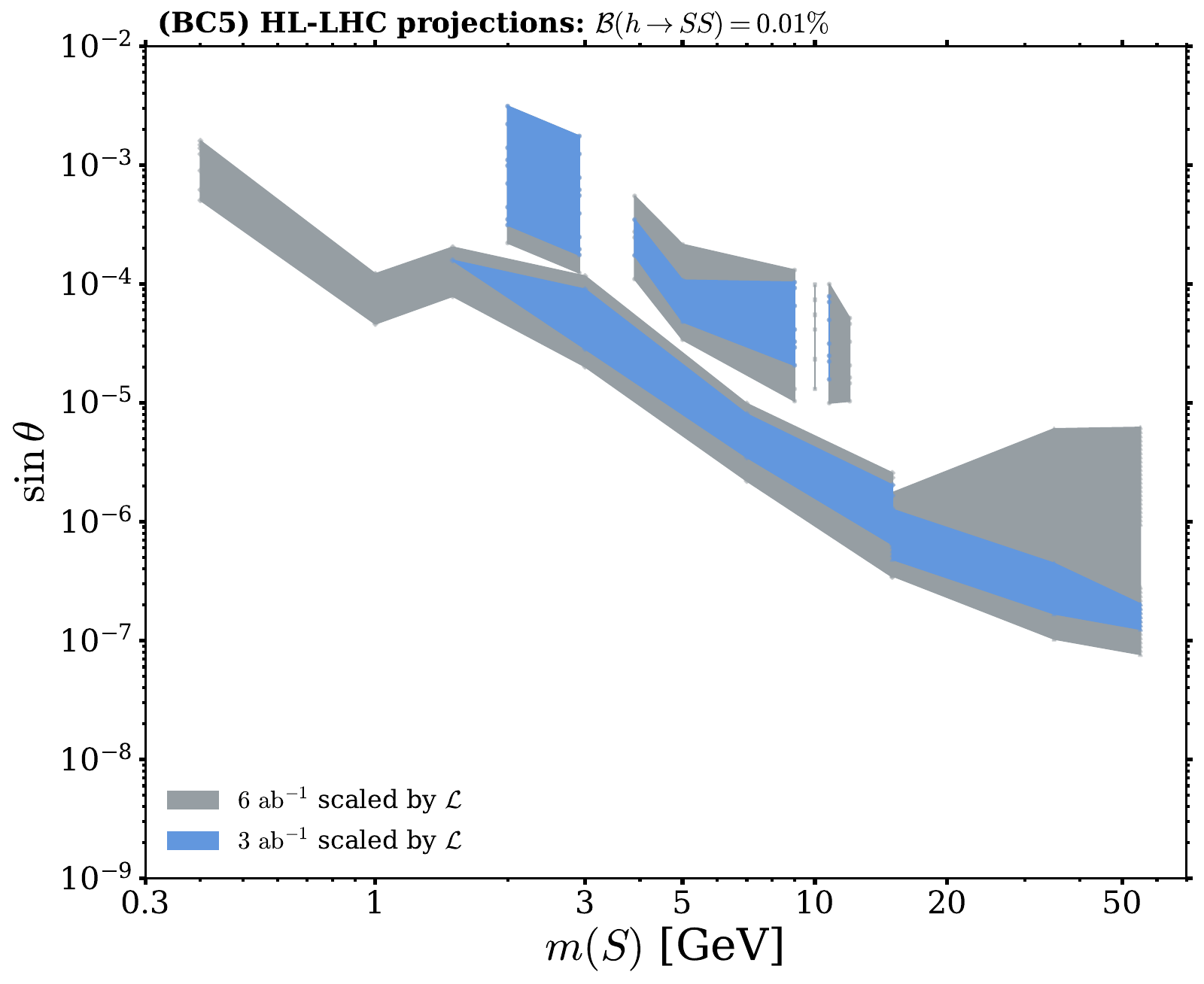}\\
  \caption{95\% CL observed upper limits in the $\{\mS, \sin \theta\}$ plane, assuming $\BhtoSS = 0.01\%$. Results are shown for HL-LHC extrapolations to 3~ab$^{-1}$ (blue) and 6~ab$^{-1}$ (gray) scaled by ${\cal L}$.}
  \label{fig:HL-LHC_0.001}
\end{figure}

Finally, Fig.~\ref{fig:HL-LHC_0.001} shows the projected sensitivity under the assumption of an ultra rare Higgs boson decay with $\BhtoSS = 0.01\%$. At present, no region of the $\{\mS, \sin \theta\}$ parameter space is excluded, and an extrapolation of current constraints using $\sqrt{\cal L}$ scaling yields no new bounds. However, if backgrounds can be controlled to negligible levels (blue), it becomes possible to probe this region, particularly in the 6~ab$^{-1}$ scenario (gray), thus fully exploiting the HL-LHC dataset. The strong dependence of the expected exclusion on the extrapolation assumptions manifests the importance of suppressing backgrounds in LLP searches, which will be one of the key challenges for the analysis of HL-LHC data.

\section{Conclusions}

In this work, we have reinterpreted selected ATLAS and CMS published results on Higgs boson decays into long-lived particles (LLPs) in the framework of the standard model extended by a light scalar, also known as the scalar portal Benchmark Case 5 (BC5) proposed by the Physics Beyond Colliders Study group. Using public data from \texttt{HEPData}, we evaluated the sensitivity of a variety of experimental strategies, ranging from reconstruction of displaced vertices to high-multiplicity hit clusters in the muon detectors, targeting different final states. Our work quantifies, for the first time, how these complementary strategies probe distinct $\mS$, and $\sin \theta$ regions of the BC5 model phase space, highlighting the importance of a diverse search program to maximize the discovery potential. Our study also pinpoints current gaps in phase space coverage left by the different experimental signatures, thereby identifying key areas for future improvement in LLP searches. Furthermore, we extrapolate current constraints to HL-LHC conditions under different assumptions and show that, in all cases, proposed LLP experiments at CERN such as SHiP, CODEX-b, MATHUSLA, and ANUBIS extend the reach in $\sin \theta$. Finally, we emphasize the crucial role of background suppression in LLP searches for fully realizing the discovery potential of the HL-LHC, which is estimated to reach $\BhtoSS \approx 0.01\%$ in the most favorable scenarios.

All figures presented in this paper, expressed as functions of both $\sin \theta$ and $\sin^{2}\theta$, are publicly available in the GitHub~\href{https://github.com/aescalante/LLP_DarkScalarReinterpretation}{\faGithub} repository~\cite{LLPPlotsRepo}.

\acknowledgments

AEV sincerely thanks Viatcheslav Valuev for his comments and suggestions on the paper, and Maksym Ovchynnikov for providing the inputs needed for the calculation of the scalar width and branching fractions. AEV also thanks Iris Capilla for her careful reading of the manuscript. The work of AEV is supported by the Atracci\'on de Talento Investigador program of the Comunidad de Madrid under project 2022-T1/TIC-23778, and by the Consolidaci\'on Investigadora program of the Agencia Estatal de Investigaci\'on through project CNS-2024-154769.


\bibliographystyle{JHEP}
\bibliography{biblio.bib}

@article{Beacham:2019nyx,
    author = "Beacham, J. and others",
    title = "{Physics Beyond Colliders at CERN: Beyond the Standard Model Working Group Report}",
    eprint = "1901.09966",
    archivePrefix = "arXiv",
    primaryClass = "hep-ex",
    reportNumber = "CERN-PBC-REPORT-2018-007",
    doi = "10.1088/1361-6471/ab4cd2",
    journal = "J. Phys. G",
    volume = "47",
    number = "1",
    pages = "010501",
    year = "2020"
}

@article{Cepeda:2019klc,
    author = "Cepeda, M. and others",
    editor = "Dainese, Andrea and Mangano, Michelangelo and Meyer, Andreas B. and Nisati, Aleandro and Salam, Gavin and Vesterinen, Mika Anton",
    title = "{Report from Working Group 2}: {Higgs Physics at the HL-LHC and HE-LHC}",
    eprint = "1902.00134",
    archivePrefix = "arXiv",
    primaryClass = "hep-ph",
    reportNumber = "CERN-LPCC-2018-04",
    doi = "10.23731/CYRM-2019-007.221",
    journal = "CERN Yellow Rep. Monogr.",
    volume = "7",
    pages = "221--584",
    year = "2019"
}

@article{Boiarska:2019jym,
    author = "Boiarska, Iryna and Bondarenko, Kyrylo and Boyarsky, Alexey and Gorkavenko, Volodymyr and Ovchynnikov, Maksym and Sokolenko, Anastasia",
    title = "{Phenomenology of GeV-scale scalar portal}",
    eprint = "1904.10447",
    archivePrefix = "arXiv",
    primaryClass = "hep-ph",
    doi = "10.1007/JHEP11(2019)162",
    journal = "JHEP",
    volume = "11",
    pages = "162",
    year = "2019"
}

@article{Bauer:2019vqk,
    author = "Bauer, Martin and Brandt, Oleg and Lee, Lawrence and Ohm, Christian",
    title = "{ANUBIS: Proposal to search for long-lived neutral particles in CERN service shafts}",
    eprint = "1909.13022",
    archivePrefix = "arXiv",
    primaryClass = "physics.ins-det",
    month = "9",
    year = "2019"
}

@article{CODEX-b:2019jve,
    author = "Aielli, Giulio and others",
    title = "{Expression of interest for the CODEX-b detector}",
    eprint = "1911.00481",
    archivePrefix = "arXiv",
    primaryClass = "hep-ex",
    doi = "10.1140/epjc/s10052-020-08711-3",
    journal = "Eur. Phys. J. C",
    volume = "80",
    number = "12",
    pages = "1177",
    year = "2020"
}

@article{Curtin:2018mvb,
    author = "Curtin, David and others",
    title = "{Long-Lived Particles at the Energy Frontier: The MATHUSLA Physics Case}",
    eprint = "1806.07396",
    archivePrefix = "arXiv",
    primaryClass = "hep-ph",
    reportNumber = "FERMILAB-PUB-18-264-T",
    doi = "10.1088/1361-6633/ab28d6",
    journal = "Rept. Prog. Phys.",
    volume = "82",
    number = "11",
    pages = "116201",
    year = "2019"
}

@article{Lee:2018pag,
    author = "Lee, Lawrence and Ohm, Christian and Soffer, Abner and Yu, Tien-Tien",
    title = "{Collider Searches for Long-Lived Particles Beyond the Standard Model}",
    eprint = "1810.12602",
    archivePrefix = "arXiv",
    primaryClass = "hep-ph",
    doi = "10.1016/j.ppnp.2019.02.006",
    journal = "Prog. Part. Nucl. Phys.",
    volume = "106",
    pages = "210--255",
    year = "2019",
    note = "[Erratum: Prog.Part.Nucl.Phys. 122, 103912 (2022)]"
}

@article{Alimena:2019zri,
    author = "Alimena, Juliette and others",
    title = "{Searching for long-lived particles beyond the Standard Model at the Large Hadron Collider}",
    eprint = "1903.04497",
    archivePrefix = "arXiv",
    primaryClass = "hep-ex",
    doi = "10.1088/1361-6471/ab4574",
    journal = "J. Phys. G",
    volume = "47",
    number = "9",
    pages = "090501",
    year = "2020"
}

@article{Strassler:2006im,
    author = "Strassler, Matthew J. and Zurek, Kathryn M.",
    title = "{Echoes of a hidden valley at hadron colliders}",
    eprint = "0604261",
    archivePrefix = "arXiv",
    doi = "10.1016/j.physletb.2007.06.055",
    journal = "Phys. Lett. B",
    volume = "651",
    pages = "374--379",
    year = "2007"
}

@article{Han:2007ae,
    author = "Han, Tao and Si, Zongguo and Zurek, Kathryn M. and Strassler, Matthew J.",
    title = "{Phenomenology of hidden valleys at hadron colliders}",
    eprint = "0712.2041",
    archivePrefix = "arXiv",
    primaryClass = "hep-ph",
    reportNumber = "MADPH-07-1502, SDU-HEP-071201, RUNHETC-2007-31",
    doi = "10.1088/1126-6708/2008/07/008",
    journal = "JHEP",
    volume = "07",
    pages = "008",
    year = "2008"
}

@article{Curtin:2014cca,
    author = "Curtin, David and Essig, Rouven and Gori, Stefania and Shelton, Jessie",
    title = "{Illuminating Dark Photons with High-Energy Colliders}",
    eprint = "1412.0018",
    archivePrefix = "arXiv",
    primaryClass = "hep-ph",
    reportNumber = "YITP-SB-14-49",
    doi = "10.1007/JHEP02(2015)157",
    journal = "JHEP",
    volume = "02",
    pages = "157",
    year = "2015"
}

@article{Carena:2022yvx,
    author = "Carena, Marcela and Kozaczuk, Jonathan and Liu, Zhen and Ou, Tong and Ramsey-Musolf, Michael J. and Shelton, Jessie and Wang, Yikun and Xie, Ke-Pan",
    title = "{Probing the Electroweak Phase Transition with Exotic Higgs Decays}",
    eprint = "2203.08206",
    archivePrefix = "arXiv",
    primaryClass = "hep-ph",
    reportNumber = "FERMILAB-CONF-22-178-T",
    doi = "10.31526/lhep.2023.432",
    journal = "LHEP",
    volume = "2023",
    pages = "432",
    year = "2023"
}

@article{CMS:2022dwd,
    author = "Tumasyan, Armen and others",
    collaboration = "CMS",
    title = "{A portrait of the Higgs boson by the CMS experiment ten years after the discovery}",
    eprint = "2207.00043",
    archivePrefix = "arXiv",
    primaryClass = "hep-ex",
    reportNumber = "CMS-HIG-22-001, CERN-EP-2022-039",
    doi = "10.1038/s41586-022-04892-x",
    journal = "Nature",
    volume = "607",
    number = "7917",
    pages = "60--68",
    year = "2022",
    note = "[Erratum: Nature 623, (2023)]"
}

@article{ATLAS:2022vkf,
    author = "Aad, Georges and others",
    collaboration = "ATLAS",
    title = "{A detailed map of Higgs boson interactions by the ATLAS experiment ten years after the discovery}",
    eprint = "2207.00092",
    archivePrefix = "arXiv",
    primaryClass = "hep-ex",
    reportNumber = "CERN-EP-2022-057",
    doi = "10.1038/s41586-022-04893-w",
    journal = "Nature",
    volume = "607",
    number = "7917",
    pages = "52--59",
    year = "2022",
    note = "[Erratum: Nature 612, E24 (2022)]"
}

@article{CMS:2021sch,
    author = "Tumasyan, Armen and others",
    collaboration = "CMS",
    title = "{Search for long-lived particles decaying into muon pairs in proton-proton collisions at $ \sqrt{s} $ = 13 TeV collected with a dedicated high-rate data stream}",
    eprint = "2112.13769",
    archivePrefix = "arXiv",
    primaryClass = "hep-ex",
    reportNumber = "CMS-EXO-20-014, CERN-EP-2021-266",
    doi = "10.1007/JHEP04(2022)062",
    journal = "JHEP",
    volume = "04",
    pages = "062",
    year = "2022"
}

@article{CMS:2024qxz,
    author = "Hayrapetyan, Aram and others",
    collaboration = "CMS",
    title = "{Search for long-lived particles decaying to final states with a pair of muons in proton-proton collisions at $ \sqrt{s} $ = 13.6 TeV}",
    eprint = "2402.14491",
    archivePrefix = "arXiv",
    primaryClass = "hep-ex",
    reportNumber = "CMS-EXO-23-014, CERN-EP-2024-025",
    doi = "10.1007/JHEP05(2024)047",
    journal = "JHEP",
    volume = "05",
    pages = "047",
    year = "2024"
}

@article{CMS:2024xzb,
    author = "Hayrapetyan, Aram and others",
    collaboration = "CMS",
    title = "{Search for light long-lived particles decaying to displaced jets in proton{\textendash}proton collisions at $\sqrt{s} = 13.6~\textrm{TeV}$}",
    eprint = "2409.10806",
    archivePrefix = "arXiv",
    primaryClass = "hep-ex",
    reportNumber = "CMS-EXO-23-013, CERN-EP-2024-225",
    doi = "10.1088/1361-6633/adaa13",
    journal = "Rept. Prog. Phys.",
    volume = "88",
    number = "3",
    pages = "037801",
    year = "2025"
}

@article{ATLAS:2024qoo,
    author = "Aad, Georges and others",
    collaboration = "ATLAS",
    title = "{Search for Light Long-Lived Particles in pp Collisions at $\sqrt{s}=13~\mathrm{TeV}$ Using Displaced Vertices in the ATLAS Inner Detector}",
    eprint = "2403.15332",
    archivePrefix = "arXiv",
    primaryClass = "hep-ex",
    reportNumber = "CERN-EP-2024-086",
    doi = "10.1103/PhysRevLett.133.161803",
    journal = "Phys. Rev. Lett.",
    volume = "133",
    number = "16",
    pages = "161803",
    year = "2024"
}

@article{ATLAS:2024ocv,
    author = "Aad, Georges and others",
    collaboration = "ATLAS",
    title = "{Search for neutral long-lived particles that decay into displaced jets in the ATLAS calorimeter in association with leptons or jets using pp collisions at $ \sqrt{\textrm{s}} $ = 13~TeV}",
    eprint = "2407.09183",
    archivePrefix = "arXiv",
    primaryClass = "hep-ex",
    reportNumber = "CERN-EP-2024-181",
    doi = "10.1007/JHEP11(2024)036",
    journal = "JHEP",
    volume = "11",
    pages = "036",
    year = "2024"
}

@article{ATLAS:2022gbw,
    author = "Aad, Georges and others",
    collaboration = "ATLAS",
    title = "{Search for events with a pair of displaced vertices from long-lived neutral particles decaying into hadronic jets in the ATLAS muon spectrometer in pp collisions at $\sqrt s$=13{\,}{\,}TeV}",
    eprint = "2203.00587",
    archivePrefix = "arXiv",
    primaryClass = "hep-ex",
    reportNumber = "CERN-EP-2021-195",
    doi = "10.1103/PhysRevD.106.032005",
    journal = "Phys. Rev. D",
    volume = "106",
    number = "3",
    pages = "032005",
    year = "2022"
}

@article{ATLAS:2018tup,
    author = "Aaboud, Morad and others",
    collaboration = "ATLAS",
    title = "{Search for long-lived particles produced in $pp$ collisions at $\sqrt{s}=13$~TeV that decay into displaced hadronic jets in the ATLAS muon spectrometer}",
    eprint = "1811.07370",
    archivePrefix = "arXiv",
    primaryClass = "hep-ex",
    reportNumber = "CERN-EP-2018-241",
    doi = "10.1103/PhysRevD.99.052005",
    journal = "Phys. Rev. D",
    volume = "99",
    number = "5",
    pages = "052005",
    year = "2019"
}

@article{CMS:2024zhe,
    author = "Hayrapetyan, Aram and others",
    collaboration = "CMS",
    title = "{Enriching the physics program of the CMS experiment via data scouting and data parking}",
    eprint = "2403.16134",
    archivePrefix = "arXiv",
    primaryClass = "hep-ex",
    reportNumber = "CMS-EXO-23-007, CERN-EP-2024-068",
    doi = "10.1016/j.physrep.2024.09.006",
    journal = "Phys. Rept.",
    volume = "1115",
    pages = "678--772",
    year = "2025"
}

@techreport{ATL-PHYS-PUB-2025-002,
    collaboration = "ATLAS",
    title         = "{Common ATLAS and CMS summary plots for Higgs boson
                    mediated hidden sectors involving long-lived particles
                    (Winter 2024-2025)}",
    institution   = "CERN",
    reportNumber  = "ATL-PHYS-PUB-2025-002",
    address       = "Geneva",
    year          = "2025",
    url           = "https://cds.cern.ch/record/2924617",
    note          = "All figures including auxiliary figures are available at
                    https://atlas.web.cern.ch/Atlas/GROUPS/PHYSICS/PUBNOTES/ATL-PHYS-PUB-2025-002",
}

@misc{CMS_Summary,
    title = "{Higgs LLP ATLAS + CMS summary plots}",
    author = "{CMS Exotica Summary plots}",
    url = "https://twiki.cern.ch/twiki/bin/view/CMSPublic/SummaryPlotsEXO13TeV#Higgs_LLP_ATLAS_CMS_summary_plot",
    howpublished = "\url{https://twiki.cern.ch/twiki/bin/view/CMSPublic/SummaryPlotsEXO13TeV#Higgs_LLP_ATLAS_CMS_summary_plot}",
    year = "2025"
    }

@article{CMS:2024bvl,
    author = "Hayrapetyan, Aram and others",
    collaboration = "CMS",
    title = "{Search for long-lived particles decaying in the CMS muon detectors in proton-proton collisions at $\sqrt{s}=13~\mathrm{TeV}$}",
    eprint = "2402.01898",
    archivePrefix = "arXiv",
    primaryClass = "hep-ex",
    reportNumber = "CMS-EXO-21-008, CERN-EP-2024-008",
    doi = "10.1103/PhysRevD.110.032007",
    journal = "Phys. Rev. D",
    volume = "110",
    number = "3",
    pages = "032007",
    year = "2024"
}

@article{PBC:2025sny,
    author = "Alemany Fern{\'a}ndez, R. and others",
    collaboration = "PBC",
    title = "{Summary Report of the Physics Beyond Colliders Study at CERN}",
    eprint = "2505.00947",
    archivePrefix = "arXiv",
    primaryClass = "hep-ex",
    reportNumber = "CERN-PBC-REPORT-2025-003",
    month = "5",
    year = "2025"
}

@techreport{CMS:2025ffw,
    collaboration = "CMS",
    title         = "{Long-lived particle triggers at CMS: Strategy and performance in proton-proton collisions at $\sqrt{s}=13.6~\mathrm{TeV}$}",
    institution   = "CERN",
    reportNumber  = "CMS-PAS-EXO-23-016",
    address       = "Geneva",
    year          = "2025",
    url           = "https://cds.cern.ch/record/2937649",
}

@article{ATLAS:2024xna,
    author = "Aad, Georges and others",
    collaboration = "ATLAS",
    title = "{The ATLAS trigger system for LHC Run 3 and trigger performance in 2022}",
    eprint = "2401.06630",
    archivePrefix = "arXiv",
    primaryClass = "hep-ex",
    reportNumber = "CERN-EP-2023-299",
    doi = "10.1088/1748-0221/19/06/P06029",
    journal = "JINST",
    volume = "19",
    number = "06",
    pages = "P06029",
    year = "2024"
}

@article{Vagnoni:2025qfv,
    author = "{LHCb Collaboration}",
    title = "{Discovery potential of LHCb Upgrade II}",
    eprint = "2503.23087",
    archivePrefix = "arXiv",
    primaryClass = "hep-ex",
    reportNumber = "LHCb-PUB-2025-001",
    month = "3",
    year = "2025"
}

@article{SHiP:2021nfo,
    author = "Ahdida, C. and others",
    collaboration = "SHiP",
    title = "{The SHiP experiment at the proposed CERN SPS Beam Dump Facility}",
    eprint = "2112.01487",
    archivePrefix = "arXiv",
    primaryClass = "physics.ins-det",
    doi = "10.1140/epjc/s10052-022-10346-5",
    journal = "Eur. Phys. J. C",
    volume = "82",
    number = "5",
    pages = "486",
    year = "2022"
}

@techreport{CMS-PAS-EXO-24-034,
    collaboration = "CMS",
    title         = "{Search for light scalar particles from Higgs boson decays
                       in exclusive final states with two muons and two hadrons}",
    institution   = "CERN",
    reportNumber  = "CMS-PAS-EXO-24-034",
    address       = "Geneva",
    year          = "2025",
    url           = "https://cds.cern.ch/record/2938044",
}

@article{ATLAS:2025pak,
    author = "Aad, Georges and others",
    collaboration = "ATLAS",
    title = "{Search for events with one displaced vertex from long-lived neutral particles decaying into hadronic jets in the ATLAS muon spectrometer in $pp$ collisions at $\sqrt{s}=13$ TeV}",
    eprint = "2503.20445",
    archivePrefix = "arXiv",
    primaryClass = "hep-ex",
    reportNumber = "CERN-EP-2025-062",
    month = "3",
    year = "2025"
}

@techreport{CMS-PAS-EXO-24-016,
    collaboration = "CMS",
    title         = "{Search for long-lived particles decaying into muons in
                    proton-proton collisions at $\sqrt{s}=13.6~\mathrm{TeV}$
                    using the CMS scouting data sets}",
    institution   = "CERN",
    reportNumber  = "CMS-PAS-EXO-24-016",
    address       = "Geneva",
    year          = "2025",
    url           = "https://cds.cern.ch/record/2938078",
}

@article{Maguire:2017ypu,
    author = "Maguire, Eamonn and Heinrich, Lukas and Watt, Graeme",
    editor = "Mount, Richard and Tull, Craig",
    title = "{HEPData: a repository for high energy physics data}",
    eprint = "1704.05473",
    archivePrefix = "arXiv",
    primaryClass = "hep-ex",
    reportNumber = "IPPP-17-31",
    doi = "10.1088/1742-6596/898/10/102006",
    journal = "J. Phys. Conf. Ser.",
    volume = "898",
    number = "10",
    pages = "102006",
    year = "2017"
}

@article{ATLAS:2024vnc,
    author = "Aad, Georges and others",
    collaboration = "ATLAS",
    title = "{Search for displaced leptons in $\sqrt{s}=13$ TeV and $13.6$ TeV $pp$ collisions with the ATLAS detector}",
    eprint = "2410.16835",
    archivePrefix = "arXiv",
    primaryClass = "hep-ex",
    reportNumber = "CERN-EP-2024-257",
    doi = "10.1103/w8hh-xf24",
    journal = "Phys. Rev. D",
    volume = "112",
    number = "1",
    pages = "012016",
    year = "2025"
}

@misc{LLPPlotsRepo,
  author       = {Escalante del Valle, Alberto},
  title        = {Supplementary material},
  howpublished = {\url{https://github.com/aescalante/LLP_DarkScalarReinterpretation}}
}

@article{OConnell:2006rsp,
    author = "O'Connell, Donal and Ramsey-Musolf, Michael J. and Wise, Mark B.",
    title = "{Minimal Extension of the Standard Model Scalar Sector}",
    eprint = "0611014",
    archivePrefix = "arXiv",
    reportNumber = "CALT-68-2614",
    doi = "10.1103/PhysRevD.75.037701",
    journal = "Phys. Rev. D",
    volume = "75",
    pages = "037701",
    year = "2007"
}

@article{Cepeda:2021rql,
    author = "Cepeda, Maria and Gori, Stefania and Outschoorn, Verena Martinez and Shelton, Jessie",
    title = "{Exotic Higgs Decays}",
    eprint = "2111.12751",
    archivePrefix = "arXiv",
    primaryClass = "hep-ph",
    doi = "10.1146/annurev-nucl-102319-024147",
    month = "11",
    year = "2021"
}

\end{document}